\documentclass[aps,prd,preprintnumbers,nofootinbib]{revtex4}
\usepackage{dcolumn}
\usepackage{lipsum}
\usepackage{graphicx,epsfig,subfigure}
\usepackage{float}
\usepackage{psfrag}
\usepackage{bm}
\usepackage{epstopdf}
\usepackage{latexsym}
\usepackage{multirow}
\usepackage{amsmath,amssymb,amsfonts}
\usepackage{enumerate}
\usepackage{hyperref}
\usepackage{multirow}
\usepackage{dcolumn}
\usepackage{array}
\usepackage{filecontents,color}
\allowdisplaybreaks

\def\beq {\begin{equation}}
	\def\eeq {\end{equation}}
\def\bea {\begin{eqnarray}}
	\def\eea {\end{eqnarray}}
\def\br{\begin{eqnarray}}
	\def\er{\end{eqnarray}}
\def\nn {\nonumber}
\def\bc {\begin{center}}
	\def\ec {\end{center}}
\def\bi {\begin{itemize}}
	\def\ei {\end{itemize}}

\def\a  {\alpha}
\def\b  {\beta}

\def\d  {\delta}

\def\e  {\epsilon}

\def\k  {\kappa}

\def\n  {\nu}
\def\o  {\omega}
\def\O  {\Omega}

\def\P  {\Pi}
\def\th {\theta}

\def\t  {\tau}

\def\pa {\partial}
\def\f {\frac}

\def\vph {\varphi}
\def\l{\left}
\def\r{\right}
\def\dis{\displaystyle}

\begin{document}
	\title{Evidence of quantum phase transition in real-space vacuum entanglement of higher derivative scalar quantum field theories}
\author{S.\ Santhosh Kumar$^{1,}$} \email[email: ]{santhu@iisertvm.ac.in} 
\author{S.\ Shankaranarayanan$^{1,2,}$} \email[email: ]{shanki@phy.iitb.ac.in}
	\affiliation{$^1$School of Physics, Indian Institute of Science Education and Research Thiruvananthapuram, CET Campus, Thiruvananthapuram 695 016, Kerala, India\\
	$^2$Department of Physics, Indian Institute of Technology Bombay, Mumbai 400 076, Maharashtra, India}
%	\affil[* ]{santhu@iisertvm.ac.in ~(Corresponding author),} 
%	\affil[+ ]{shanki@phy.iitb.ac.in}	
	\begin{abstract}
In a bipartite set-up, the vacuum state of a free Bosonic scalar field
is entangled in real space and satisfies the area-law---
entanglement entropy scales linearly with area of the boundary between
the two partitions. In this work, we show that the area law is
violated in two spatial dimensional model Hamiltonian having dynamical critical exponent 
$z=3$. The model physically corresponds to
next-to-next-to-next nearest neighbour coupling terms on a
lattice. The result reported here is the first of its kind of
violation of area law in Bosonic systems in higher dimensions and
signals the evidence of a quantum phase transition. We provide
evidence for quantum phase transition both numerically and
analytically using quantum Information tools like entanglement
spectra, quantum fidelity, and gap in the energy spectra. We identify
the cause for this transition due to the accumulation of large number
of angular zero modes around the critical point which catalyses the
change in the ground state wave function due to the
next-to-next-to-next nearest neighbor coupling. Lastly, using
Hubbard-Stratanovich transformation, we show that the effective
Bosonic Hamiltonian can be obtained from an interacting fermionic
theory and provide possible implications for condensed matter systems.
\end{abstract}
\maketitle
	\section{Introduction}
Quantum field theory plays a crucial role in understanding some of the
interesting features of low-temperature condensed matter systems
\cite{wen2004quantum,fradkinbook}. More importantly, it is well
studied in the case of quantum phase transitions (QPTs)--- transitions
at absolute zero --- where the abrupt change in the ground state
properties of many body systems is governed by coherent quantum
fluctuations resulting from the Heisenberg uncertainty principle
\cite{hertz1976-PRB,sondhi1997rmp,Senthil2004-science,Vojta2000,Osterloh2002,Jacob2002,Vojta2003,
  vidal2003,vedral2008,subir,carr2011,batrouni}.  The transition from
one quantum phase to another is brought about due to the changes in
the external parameter $(P)$ of the system described by the
Hamiltonian $H(P)$.  As like classical phase transition, at quantum
critical point (QCP), there are long-range correlations in the system
\cite{Vojta2003,subir,carr2011} and that the state of the system is
strongly entangled \cite{vedral2008,cardy2006}. Hence, it is expected
that quantum entanglement across regions may play a crucial role at
QCP \cite{Osterloh2002,Nielsen2002,vidal2003,Larsson2005}.

In order to overcome the complexity of the interactions underlying
QPTs, theoretical work has focused on one-dimensional quantum systems
\cite{subir,carr2011,vedral2008,Giamarchi}.  In this work, we study
the effect of next-to-next-to-next nearest neighbor (NNN) coupling
terms on the quantum fluctuations. More specifically, we consider a two-dimensional model that corresponds to NNN coupling terms on a
lattice. Interesting features of the model are as follows: (i) It
corresponds to the Hamiltonian of a system of coupled harmonic
oscillator and hence one can compute all the relevant quantities
analytically.  (ii) The celebrated entanglement entropy-area law is
valid for systems having local interactions~\cite{unanyan2005,
  Barthel2006, eisert2007}. We show violation of area law
due to the presence of NNN coupling terms on a lattice. (iii) The
dynamical critical exponent --- a quantity that measures the scaling
anisotropy between space-time variables --- for our model is
$z=3$. This needs to be contrasted with the quantum Lifshitz
transitions \cite {revaz1999prb,fradkin2004ap,fradkinbook} where the
dynamical critical exponent is $z=2$ and QCP is conformally critical
\cite{fradkin2006prl,fradkin2009jpa}.
				
Although there is yet no fundamental understanding on the role of
quantum correlations, there is a huge body of work to identify
quantum information theoretic tools indicating quantum criticality
\cite{pcoleman2005,subir2008}:
	%besides entanglement entropy, there have other quantifying
	%tools indicating the on-set of QPTs Fidelity measures
	%similarity between two states~\cite{gu2010} (i) At QCP,
quantum fidelity drastically changes due to the external parameter
$(P)$ thus signifying the notion of quantum order
parameter~\cite{gu2010}.
	% (ii) Given a pure bipartite many particle ground state 
	$\psi_{\rm GS} (P) \in \cal{H}$ (Hilbert space), the associated reduced density matrix of its subsystem can be written as 
	$\rho_{\rm red}= \exp\left( - h_{\rm E} \right)$ \cite{Peschel2012}, where $h_{\rm E}$
is the entanglement Hamiltonian of any one of the subsystem. It has
been shown that entanglement spectrum contains more information than
the entanglement entropy~\cite{haldane2008,chandran2014}. Entanglement
Hamiltonian is well studied in the case of fractional quantum hall
states \cite{haldane2008}, one dimensional quantum spin systems
\cite{chiara2012,LAFLORENCIE2016}. In particular, at QCP, there is a
finite `energy gap' of $h_{\rm E}$
\cite{chiara2012,illuminati2013,Thomale2010(3)}. This gap is generally
termed as the entanglement gap or Schmidt gap which is considered as
an order parameter to diagnose QPTs \cite{chiara2012,LAFLORENCIE2016}.
More importantly, the fingerprints of the topological order is encoded
in the entanglement spectra
\cite{haldane2008,pollmann2010prb,Thomale2010(2),martin2016prb}.
	
The rest of the article is organised in the following manner: In
Sec. (\ref{sec1}), we discuss the model Hamiltonian and the method for
studying the entanglement properties. It also contains details
about mapping the model Hamiltonian to a system of coupled harmonic
oscillators. We also briefly discuss the method to evaluate
entanglement entropy (EE) from a pure Gaussian ground state~\cite{srednicki93}.
Sec. (\ref{sec2}) contains the details of the numerical and analytical
tools in support of the observed QPT in the model. Finally,
Sec. (\ref{sec3}) concludes with the important outlook of the model
Hamiltonian in strongly interacting systems. It also comprises of the
details about the calculation of our model Hamiltonian from an
interacting fermionic theory and discuss possible applications to 
condensed matter systems.

	\section{ Model and Setup}
In this section, we discuss the model Hamiltonian and provide
essential steps to obtain entanglement entropy. We also show how
our model Hamiltonian can be mapped to a system of coupled harmonic
oscillators. Finally, we
discuss the real-time method to evaluate the ground state reduced
density matrix for the model Hamiltonian and the approach to evaluate
the ground state entanglement entropy.
%	\label{sec1}	
\subsection{Model Hamiltonian}
\label{sec1}
We consider  the following two dimensional Hamiltonian:
\beq
\label{h2}
\dis	H =\frac{1}{2}\int d^2 {\bf r}\left[|\hat \Pi({\bf r})|^2 + 
|\nabla \hat\Phi({\bf r})|^2+ \frac{\epsilon}{\kappa^2}|\nabla^2 \hat\Phi({\bf r})|^2 
+\frac{\tau}{\kappa^4}  |\nabla^3\hat\Phi({\bf r})|^2   \r] ,
\eeq
\noindent where $\hat\P({\bf r})$ is the conjugate momenta of
the massless Bosonic scalar field $\hat\Phi({\bf r})$,
$\epsilon$ and $\tau$ are dimensionless constants (and here it
take values either $0$ or $1$), $\kappa$ has the dimension of
wave number which sets the scale for the deviation from
Lorentz invariance and the speed of propagation is set to
unity, $\nabla^2 $ is the two dimensional Laplace operator,
$\nabla^4$ and $\nabla^6$ are the higher order spatial
operators of the circular coordinates ${\bf r} (=r,\th) $ and
$c=\hbar = 1$.
The equal time canonical  commutation relation between fields  is given by,
\beq
\left[\hat\Phi({\bf r}),\hat\Pi({\bf r'} )\right] = i\,\delta^2 ({\bf r-r'})= \frac{i}{r}\delta ( r-r')\delta ({ \th-\th'})
\eeq
Before going to the details, we briefly discuss	some of the salient features of the model Hamiltonian in Eq. (\ref{h2}): 
(i) The above Hamiltonian corresponds to free field with a non-linear
dispersion relation between frequency $\omega$ and wave
number $k$ via, $\o^2= k^2+\e\,k^4/\k^2+\t\,k^6/\k^4$. 
(ii) It is well-known that $\nabla^4$ term with non-linear terms lead
to classical Lifshitz transitions
\cite{hornreich75,Hornreich1985prb}. Here, we show that
$\nabla^6$ terms or NNN coupling drives QPT
~\cite{shanki2012}.
(iii) It is important to note the differences between
quantum Lifshitz transitions and our case. In the quantum
Lifshitz transition, scalar field theory has dynamical
critical exponent $z=2$ and the dispersion relation is $\o^2=
k^2+\e\,k^4/\k^2$
\cite{fradkin2004ap,fradkinbook,subir,Ardonne2004}. However,
in our case, the dynamical critical exponent is $z=3$ and
$\o^2= k^2+\e\,k^4/\k^2+\t\,k^6/\k^4$ is the dispersion
relation.
%In the case of Lifshitz transition, the dynamical exponent is $z=2$~\cite{subir,Ardonne2004}. However, in our case the dynamical exponent is   $z=3$.  		
%	
(iv) It is known that quantum fluctuations play an important role in
lower dimensions, and almost invariably, destroy long-range
order. In the classical systems, Mermin-Wagner Theorem
\cite{coleman1973,Klein1981,Nolting2001} precludes true
long-range order in the thermodynamic limit at non-zero
temperature in one and two dimensions. We show that, in two
dimensions, NNN coupling introduces instability and drive
quantum fluctuations leading to QPT.

%In the rest of the section, we provide key steps to compute
%entanglement entropy from Hamiltonian (\ref{h2}) in a bipartite
%setup. In order to do that we first map our model Hamiltonian to a
%system of coupled harmonic oscillators via eigenfunction expansion in
%two dimensional space followed by a central discretization scheme
%\cite{pletcher1997computational} for all higher spatial derivatives.

\subsection{Mapping of higher order spatial Hamiltonian to a system of coupled harmonic oscillators } 	
Using the following ansatz:
	\begin{subequations}	
\br
\hat\Pi( {\bf r})= \sum_{m=-\infty}^\infty \frac{\hat\Pi_m(r)}{\sqrt{\pi r}}\cos m \theta ,\\ 
\hat\Phi({\bf r})= \sum_{m=-\infty}^\infty \frac{\hat\vph_m(r)}{\sqrt{\pi r}}\cos m \theta,
\label{h}
\er
	\end{subequations}	
\noindent we expand the two dimensional real scalar fields in
Hamiltonian (\ref{h2}) (it is important to note that the Hamiltonian
has same set of eigenvalues if we expand in sine function. For the
case of complex scalar fields, one has to expand it in terms of
exponential function). In the above expressions (\ref{h}), $m$ is the
angular momentum quantum number, $r$ and $\theta$ are the radial and
angular coordinates, respectively.

The canonical commutation relation between the  new rescaled fields is 
%fields is  between the components of the Hermitian fields in Eq:(\ref{h}) is
\beq
\label{c1}
\left[\hat \vph_m({ r}),\hat \Pi_{m'}({ r'} )\right] = i \, \delta ({ r-r'})\thinspace \delta_{mm'}
\eeq

\begin{figure}[h]
	\centering
	\includegraphics[scale=0.43]{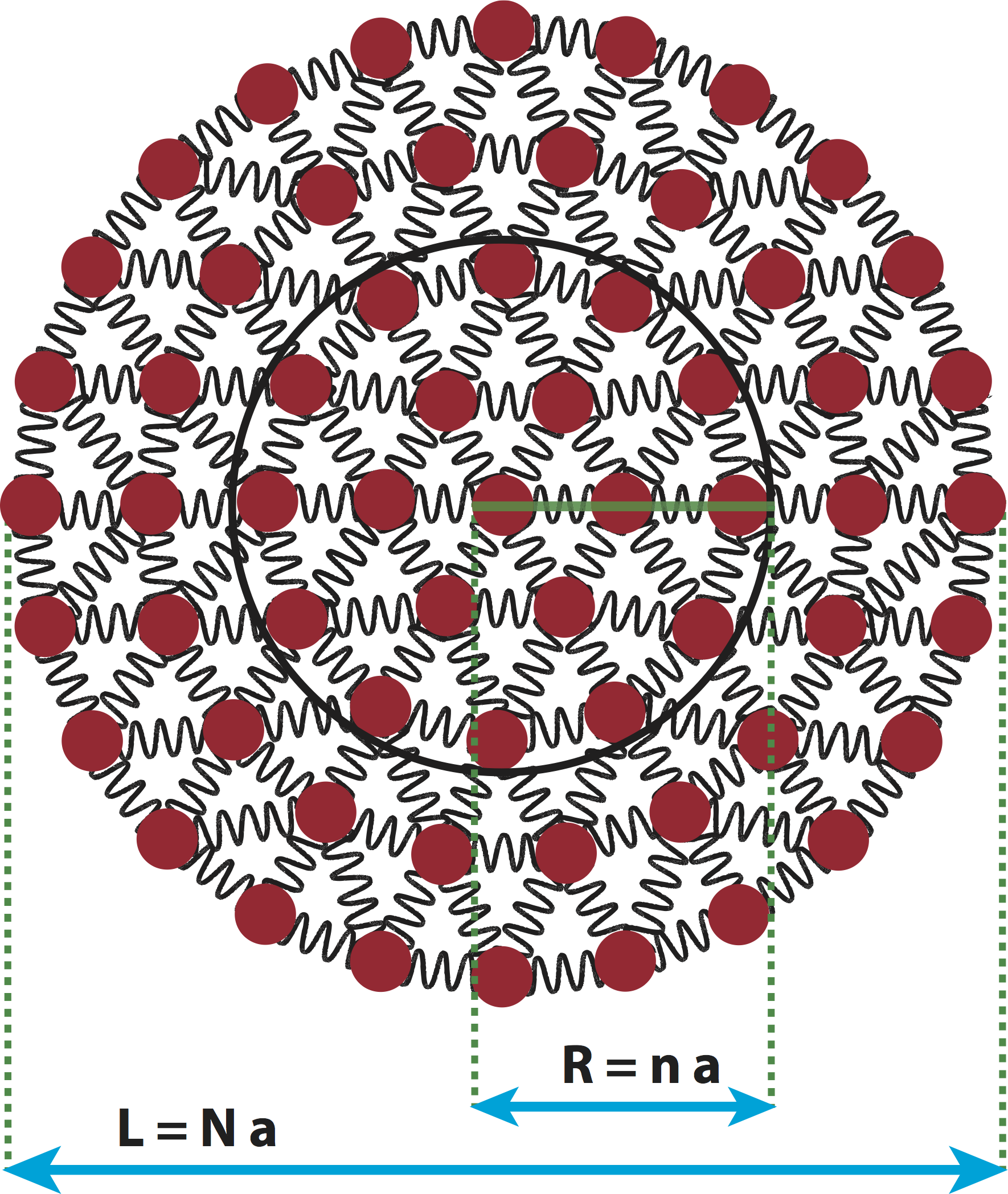}
	\caption{{A pictorial representation of the radial lattice chain. The brown dots are $N$ harmonic lattice sites arranged radially, 
	$a$ is the lattice parameter which acts as an ultraviolet cut-off and $L=Na$ is the total length of the system which act as an infrared cut-off.  
	In this arrangement, we trace over first $(N-n)$ oscillators and evaluate reduced density matrix $\rho_{\rm {red}}$ and  EE. For the nearest-neighbour and NN coupling, it can be seen that entropy scales as area of the boundary $n$, however, in the case of NNN coupling entropy scales non-trivially with $n$. }}
	\label{lattice}
\end{figure}

\noindent To map the model Hamiltonian (\ref{h2}) to a system of
coupled harmonic oscillators; we perform integration over the angular
coordinate $\theta$ by invoking orthogonal properties of cosine
functions. We then discretize all radial derivative terms using
central difference discretization scheme
\cite{pletcher1997computational}.
%It is given as,
%\begin{subequations}
%	\br
%	f'(x)&= &\frac{f(x+h)-f(x-h)}{2 h}+ O(h^2)\\
%	f''(x)&=&\frac{f(x+h)-2 f(x)+f(x-h)}{h^2}+O(h^2)\\
%	f'''(x)&=&\frac{f(x+2 h)-2 f(x+h)+2 f(x-h)+f(x-2 h)}{2 h^3}+O(h^2)
%	\er
%\end{subequations}
More explicitly, we use infrared cut-off as $L= N a $, where $N$ is
the number of Lattice points and $a$ is the ultraviolet cut-off. In
the discretized version, the dimensionless scalar fields are
expressed as $\hat\pi_{m,j}= a\, \hat\Pi_m (j a)$,
$\hat{\vph}_{m,j}=\hat\vph_m(ja)$, and radial distance is $r=ja$.
Hence, the form of Hamiltonian and commutation relation in
Eqs. (\ref{h2}) and (\ref{c1}), become (for more details see
supplementary material),

\beq
\label{eq:disc-Ham}
H(P) =
\frac{1}{2a} \sum_{m=-\infty}^\infty\sum_{i,j=1}^N
%\sum_{\substack {i,j=1,\\ m=-\infty}}^{N,\infty}
\l[\hat\pi_{m,i}^2  \delta_{ij} + \hat\vph_{m,i} \,  K_{ij} (P,m)  \,\hat\vph_{m,j} \r]
\eeq
\br \l[\hat\vph_{m,i},\hat\pi_{m',j}\r] &=& i \d_{mm'}\d_{ij} \\
\dis \mbox{where} ~~~P &=& \frac{1}{(\k a)^2} 
\er
is the coupling parameter that determines the extend
of deviation from the linear dispersion relation, and
$K_{ij}$ is the real symmetric matrix which contains
all the coupling terms--- the nearest, the
next-to-next nearest (NN), and NNN coupling, more
specifically it includes the coefficients of $P$ and
$P^2$ (see supplementary material for more details
about elements of the $K_{ij}$ matrix). As can be
seen, the Hamiltonian (\ref{eq:disc-Ham}) corresponds
to N-coupled harmonic oscillators. We use an open
boundary condition, $\hat\vph_{m,N+1}=0$ to calculate
EE and other relevant quantities.
Fig.~(\ref{lattice}) provides a pictorial representation of the radial harmonic lattice. 
%The dispersion relation corresponding to the Hamiltonian (\ref{eq:disc-Ham}) is given below:
% 
%         \begin{eqnarray}
%        a^2\omega^2(\theta)&=&a^2\left[\omega_1^2(\theta)+\epsilon P\omega_2^2(\theta)
%        +\tau P^2\omega_3^2(\theta)\right]=a^2\left[2-2\cos(4\theta)\right]+a^2\epsilon P\left[6-8\cos(2\theta)+2\cos(4\theta)\right]\nonumber\\
%        &&+a^2\tau P^2\left[10-8\cos(2\theta)-8\cos(4\theta)+8\cos(6\theta)-2\cos(8\theta)\right]    \label{diseq}    \end{eqnarray}
%        %
%        where $\dis\theta=\pi \zeta/N,\quad \zeta\in\{0,1,\dots N-1\}, \omega_1, \omega_2$ and $\omega_2$ are the normal mode frequencies of the discretized $|\nabla\phi|^2$, $|\nabla^2\phi|^2$ and $|\nabla^3\phi|^2$ terms respectively. 
The EE of the $(N-n)$ traced oscillators is given by
\cite{von1955mathematical}, \beq S=-\mbox{Tr}\l(\rho_{\rm
	red}\,\log \rho_{\rm red}\r). \eeq where $\rho_{\rm red}$
is the reduced density matrix obtained by integrating over
the $(N-n)$ oscillators. The entanglement spectra of the
entanglement Hamiltonian $h_{\rm E}$ is given
by\cite{Peschel2012}, \beq h_{\rm E}=-\log \rho_{\rm red}.
\eeq
\subsection{Evaluation of the entanglement entropy for the ground state}

As mentioned earlier, reduced density matrix ($\rho_{red})$ provides
information about the strength of the quantum correlations across
different regions.  $\rho_{red}$ is evaluated by tracing out the
quantum degrees of freedom associated with the scalar order parameter
inside a two dimensional region of radius $R$.  $\rho_{\rm red}$ can
be computed semi-analytically from Hamiltonian
(\ref{eq:disc-Ham}). 

The procedure to obtain the entanglement entropy is similar to
the one discussed in Refs.~\cite{srednicki93,shanki2012}. We
assume that the quantum state corresponding to the Hamiltonian
of the $N$-harmonic oscillator system is the ground state
i. e. $\psi_{\rm GS} (x_1, x_2,\cdots , x_n, t_1,t_2, \cdots,
t_{N - n})$.  $\rho_{red}$ is obtained for $n$ oscillators by
tracing over $(N-n)$ of the $N$ oscillators. von Neumann
entropy quantifies the ground state entanglement of a
bipartite system via $S=-\mbox{Tr}\l(\rho_{\rm red}\,\log
\rho_{\rm red}\r)$ \cite{bombelli86,srednicki93}.

The ground state wave function of the above Hamiltonian in Eq.  (\ref{eq:disc-Ham}) is given by,
\beq
\psi_{\rm GS}(X)=\l( \frac{Det~ \Lambda}{\pi^{N}}\r)^{1/4} \exp{(-X^T.\Lambda. \,X/2)}
\eeq 
where

$\Lambda=\sqrt{K}=\dis\begin{pmatrix}
A & B\\
B^{T} & C
\end{pmatrix} $
and  A, B, C are the sub matrices of $\Lambda$ matrix with dimensions
$ n\times n ,n\times (N-n),$ and $(N-n)\times(N-n)$ respectively such that  $N>n$ and $X=(x_1,x_2, \ldots x_n, t_1, t_2, \ldots t_{N-n})^T$. 
The ground state density matrix is 
\beq 
\rho_{\rm GS}(X,X')= \psi_{\rm GS}(X)  \; \psi^*_{\rm GS}(X')= \psi_{\rm GS} (x_1,x_2, \cdots , x_n, t_1,t_2, \cdots, t_{N - n})\,\psi^*_{\rm GS} (x_1, \cdots , x_n, t_1,t_2, \cdots, t_{N - n})
\eeq

\noindent The reduced density matrix ($\rho_{\rm red}$) for the ground state,
is obtained by tracing over the first $(N-n) $
of $N$ oscillators of the pure density matrix $\rho_{\rm GS}$
\bea
\label{d} 
\rho_{\rm red}(x,x')& =& \int d t \, \rho_{GS} (X, X')\\  &=&\int\l(\prod_{i=1}^{N-n} dt_i\r) \;   \psi_{\rm GS}(x_1,x_2, \ldots x_n, t_1, t_2, \ldots t_{N-n}) \psi^*_{\rm GS}(x_1',x_2', \ldots x_n', t_1, t_2, \ldots t_{N-n})
\eea
where $x=(x_1, x_2, \ldots, x_{n})^T$ and $t=(t_1, t_2, \ldots, t_{n})^T$. The integral in Eq.~(\ref{d}) can be evaluated explicitly and can be written as, 
\beq 
\rho_{_{\rm red}}(x,x')\sim \exp\left[ -(x^T.\Gamma.x + x'^T.\Gamma. x')/2  + x^T.\O. x'\right] 
\eeq 
where $\O$ and $\Gamma$ are defined as 
\begin{subequations}
	\label{b}
	\br
	\O&= &\frac{1}{2} B^T A^{-1} B,  \\
	\Gamma&=& C-\O
	\er
\end{subequations}
Let $\Gamma_D$ and V be the diagonal and orthogonal matrices
of $\Gamma$. Performing the transformation $x=V^T \Gamma_D
^{-1/2} y$ then $\O\rightarrow\O'=\Gamma_D^{-1/2} V \O V^T
\Gamma_D^{-1/2} $, the reduced density matrix becomes, \beq
\label{equred1}
\rho_{_{\rm red}}(y,y')\sim \exp\left[ -(y.y + y'.y')/2  + y^T.\O'. y'\right] 
\eeq 
Rewriting $y=W z$, where W is an orthogonal matrix such that $\O'$ is diagonal in W basis,  Eq.~(\ref{equred1}) becomes,
\beq \rho_{_{\rm red}}(z,z')\sim \prod_{i=1}^{n}\exp\left[ -(z_i^2+ z_i'^2)/2  + \O_i'  z_i  z_i'\right]  \eeq 
where $\O_i'$ is an eigenvalue of $\O'$. von Neumann entropy of the reduced density matrix ($\rho_{red}$) is given by,
\beq S_m =\sum_{i=1}^{n} S_{m,i} (\xi_{i}) \eeq
where
\begin{subequations}
	\br
	S_{m,i}(\xi)&=& \log{(1-\xi_{i})} -\frac{\xi_{i}}{1-\xi_{i}} \log \xi_{i},\\
	\mbox{and}~~~~~~	\xi_{i}& =& \frac{\O_{i}'}{1+\sqrt{1-\O_{i}'^2}} 
	\er
\end{subequations}
The entanglement  entropy, $S$,   is computed by summing
% In order to  compute the total entropy of the system in Eq;(\ref{hmm}), we have to sum
over all  $m$ modes as
\beq S= S_{m=0} + 2 \sum_{m=1}^{\infty} S_{m} \eeq
where  $S_{m=0}$ is the value of EE for $m=0$ and all other $S_m $ values are multiplied by  a degeneracy factor 2.
\section{Results}
\label{sec2}
In this section, we evaluate the physical quantities for the model
Hamiltonian as discussed in Sec. (\ref{sec1}).
In the first part, we obtain 
the numerical tools ---- entanglement entropy and
entanglement spectrum --- to confirm the evidence
of QPT.  In the second part, we use analytical
tools like quantum ground state fidelity and
gap in the energy spectra to identify the
cause of QPT.
\subsection{Numerically evaluated tools}
%	\subsubsection{ Importance of sixth order derivative term}
% \noindent {\it Importance of sixth order derivative term:}
We compute the entanglement entropy numerically for
the discretized Hamiltonian presented in
Eq. (\ref{eq:disc-Ham}). The computations are done
using Matlab for the lattice size $N=600$, $10 \leq n
\leq 590$ and the error in computation is
$10^{-8}$. More specifically, we choose the following
cases for the numerical evaluation of two dimensional
real-space entanglement entropy;
\begin{itemize}
	\item [(I)] $\tau =0$, $\epsilon = 1$.--- The $K_{ij}$
	matrix contains both nearest neighbour and NN
	coupling % in the $K_{ij}$ matrix. % and the
	corresponding dispersion relation is $\omega^2 = k^2
	+ k^4/\kappa^2.$
	
	\item[(II)] $\tau =1$, $\epsilon = 0$.--- The $K_{ij}$
	matrix contains nearest neighbour, NN, and NNN
	coupling and the corresponding dispersion relation
	is $\omega^2 = k^2 + k^6/\kappa^4.$
	
	\item [(III)]$\tau = 1$, $\epsilon = \pm 1$.--- The
	$K_{ij}$ matrix contains nearest neighbour, NN, and
	NNN coupling and the corresponding dispersion
	relation is $\omega^2 = k^2 \pm k^4/\kappa^2 +
	k^6/\kappa^4.$
\end{itemize}

For case (I), like in canonical scalar field, the
entanglement entropy scales linearly as $n$ for all values
of $P$.  Hence, we will not present the results in the main
text (see supplementary material for more details).  The
results for case (III) are similar to case (II), hence, we
only present the results for case (II).
\begin{figure}
	\centering
	\includegraphics[scale=.35]{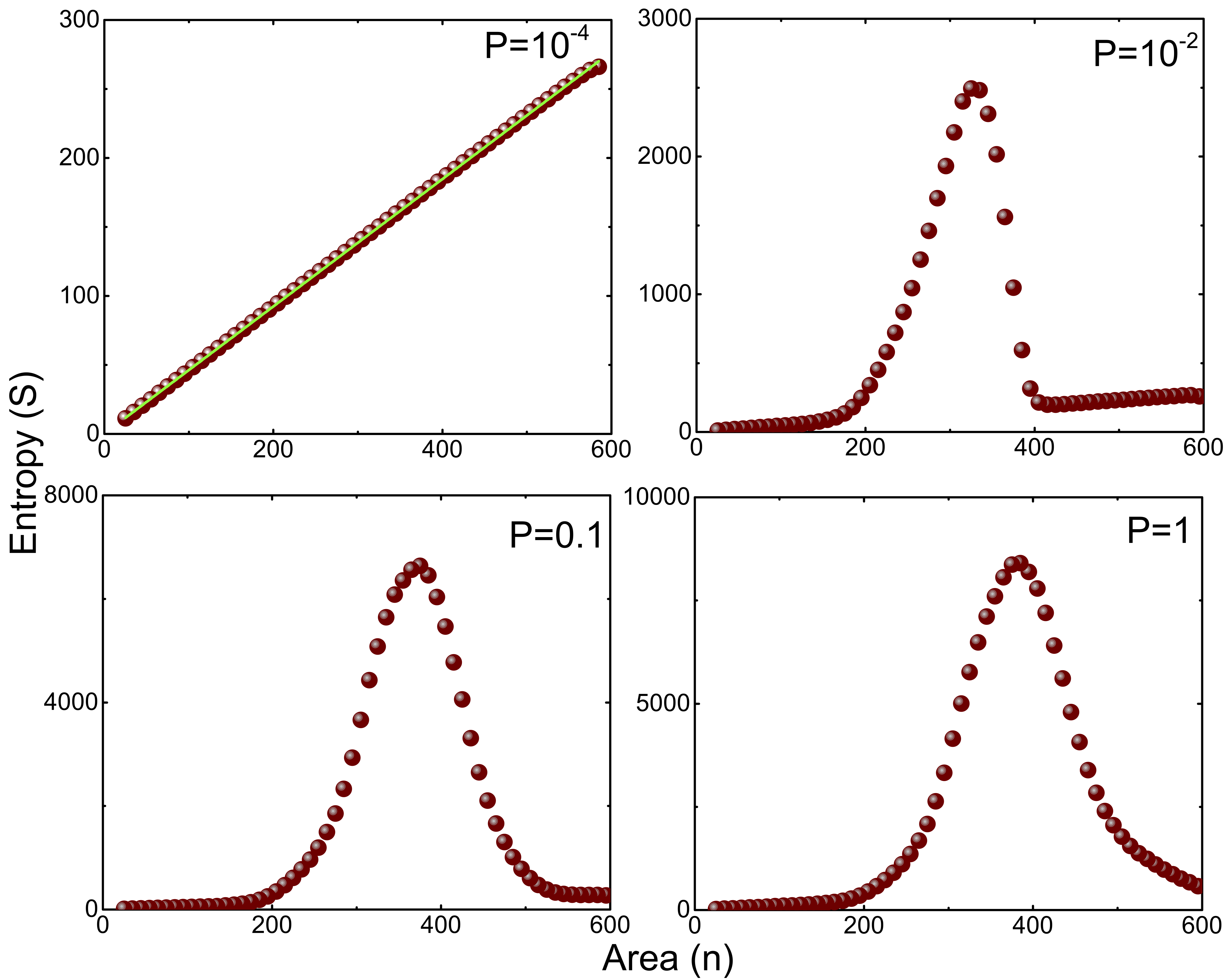}
	\caption{ Plots of two dimensional vacuum entanglement
		entropy ($S$) versus area ($n$) for case II
		($\tau =1$, $\epsilon = 0$) is shown for different
		values of $P$.  The brown dots are the numerical
		output data points and the green line is the best
		linear fit.}
	\label{fig:6thDerivative}
\end{figure}

In Fig.~(\ref{fig:6thDerivative}), we have plotted $S$ versus
$n$ for different values of the coupling constant $P$. The
following points are interesting to note: First, the entropy
profile changes as the coupling strength $P$ is increased.  In
the case of $P = 10^{-4}$, the entropy is linearly related to
$n$, however, as $P$ increases the entropy-area linear
relation is broken and the entropy changes by an order. It is
interesting to see in the case of higher values of $P$, for
large values of $n$ the entropy scales linearly with area up
to certain value of $n$ and it is highly non-linear and for
small values of $n$ the entropy scales linearly with
area. Second, it is known that the area-law is valid for
gapped systems~\cite{eisert2007,lai2013}. It is important to
note that the addition of NNN coupling make the system gap-less
and the area-law is violated.  Third, the area-law violation
presented here is a generic feature for the Hamiltonian
(\ref{eq:disc-Ham}) containing NNN coupling and it is
intrinsic to gap-less quantum Hamiltonian. In other words, the
area-law violation exists for all values of lattice points
$N$'s (see supplementary material for more details). %
Fourth, the area-law is satisfied for case (I),
dispersion relation is $\o^2= k^2+k^4/\k^2$ and dynamical
critical exponent is $z=2$, which is consistent with the
analysis reported in Ref. \cite{fradkin2006prl}. The
expression for entanglement entropy is $S= a \,n +b \,\log n$,
where $a$ and $b$ are some arbitrary constants depends on the
nature of partition used in the bipartite set-up (for more
details see figs. (1) and (2) in supplementary material and in
our analysis, $a= 1.39$ and $b=0.038$).

The model Hamiltonian in Eq. (\ref{h2}) was first studied in
three dimensional space and concluded with the following
salient remarks \cite{shanki2012}: (i) EE violates area-law in three dimensional
space. (ii) NNN coupling term is responsible for the change in
the behaviour of EE and thus changes the ground system
properties of the system at QCP. (iii) Like in the two
dimensional case, entanglement entropy in three dimensions scales
linearly with $n$ for small values of $P$. However, for large values of $P$, it scales inversely. 
In both cases, the study of EE reveals that the violation
of area-law is an inherent property of the Hamiltonian in
Eq. (\ref{eq:disc-Ham}) with NNN coupling. It is possible to
identify the following scaling transformations $ r\to r
e^{-\iota}, \hat \Phi\to \Phi e^{\iota}, \k\to \k e^{\iota} $,
where $\iota$ is the scaling parameter such that Hamiltonian
remains invariant. It is interesting to note that we have
evaluated EE in one dimension where the entropy remains a
constant, that is, area-law is followed. Hence, two dimension
is the critical dimension for observing the change in
the behaviour of EE.

The above comparison and results conclusively show that NNN
coupling lead to different quantum phases. To firmly identify
what causes the change in the entropy by an order, in the rest
of this work, we use three quantifying measures ---
entanglement spectrum, ground state fidelity, and many-body
ground state energy --- and show that this is indeed QPT. It
is important to note that the first measure is numerical while
the other two measures are analytical.
\subsubsection{Spectra of entanglement Hamiltonian}
%\noindent {\it Spectra of entanglement Hamiltonian:} 
Reduced density matrix contains complete information about quantum
entanglement, however, entanglement entropy being scalar may
not provide complete information \cite{haldane2008}.  Entanglement 
Hamiltonian is an imaginary system that describes
the correlations of the ground state.  While it cannot be
measured directly, it is related to the statistics of the
fluctuations in the lattice. Entanglement Hamiltonian $(h_{\rm
	E})$ is defined as $h_E = - \log \rho_{\rm red}$ and plays
the role of $\beta H$ in thermodynamic systems
\cite{haldane2008}.  It has been shown that the largest and
the second largest eigenvalues of $h_{\rm E}$ forms a gap at
the QCP \cite{haldane2008} and widely considered as a tool for
quantifying QCP
\cite{chiara2012,illuminati2013,LAFLORENCIE2016}. In the same
spirit, we evaluate the first two largest eigenvalues of
$h_{\rm E}$ and verify the formation of gap at QCP.

\begin{figure}
	\includegraphics[scale=0.41]{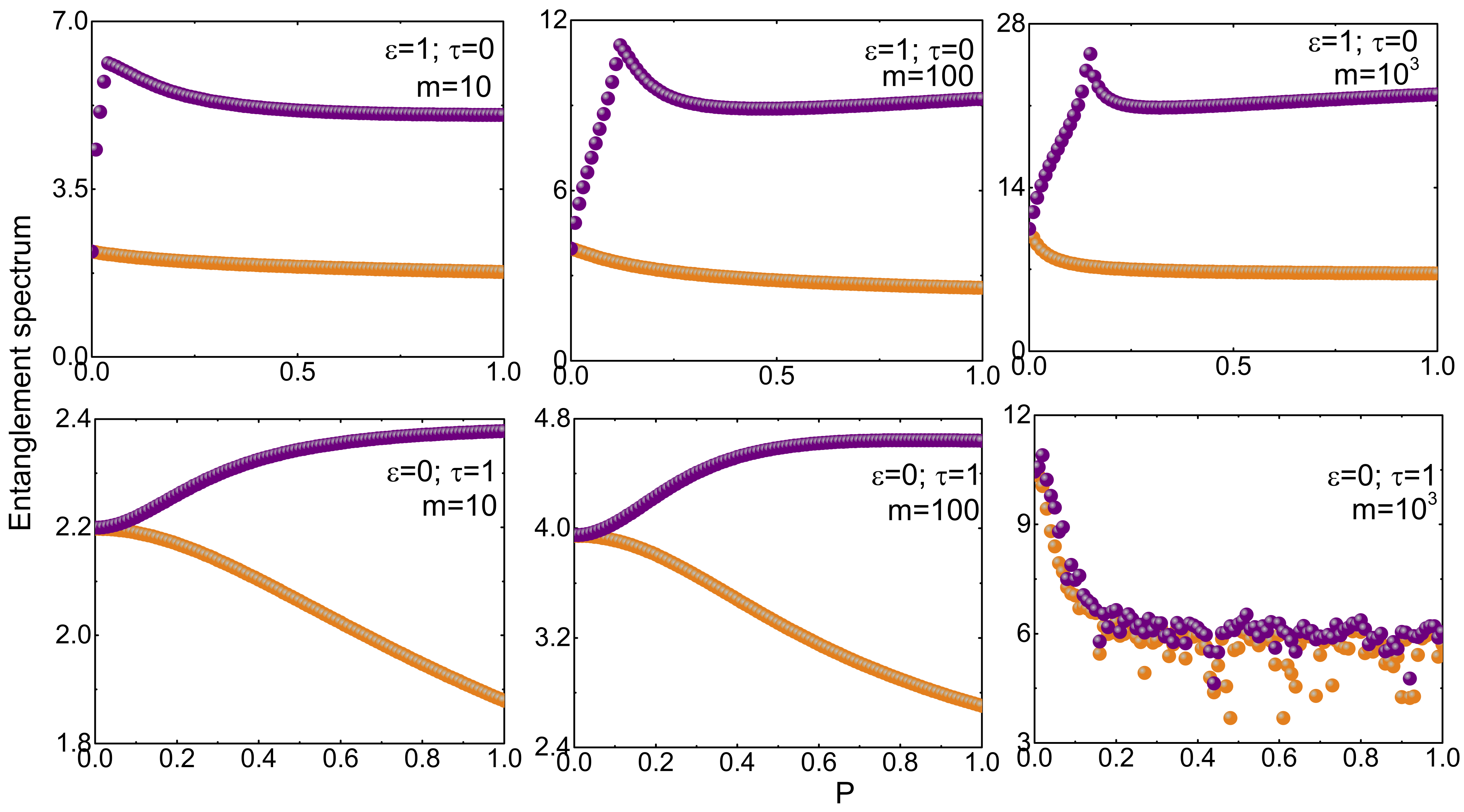}
	\centering
	\caption{ Plots show the variation of the
		largest (orange) and second largest (purple)
		eigenvalues of the entanglement spectrum as
		a function $P$ for different $m$ values. The
		top panel is for case I ($\epsilon = 1, \tau
		= 0, n=300$) while the bottom panel is for
		case II ($\epsilon = 0, \tau = 1, n=300$). }
	\label{Fig:entangspe}
\end{figure}

In Fig. (\ref{Fig:entangspe}), largest and the second largest
eigenvalue of $h_{\rm E}$ are plotted for different coupling
parameter $P$ for NN and NNN coupling, from which we infer the
following: (i) In the case of NN coupling, the largest and the
second largest eigenvalues of $h_{\rm E}$ have a gap for all
values of the coupling constant except at $P = 0$. (ii) In the
case of NNN coupling, the largest and the second largest
eigenvalues of $h_{\rm E}$ is degenerate for $P < 0.17$,
however, above the critical point $P = 0.17$ the two
eigenvalues are non-degenerate.  (iii) The presence of the
entanglement gap at a finite $P$ provides an evidence of
quantum critical point at $P = 0.17$. It is also interesting
to note that it is at the same point that the overlap function
also shows a sharp change \cite{illuminati2013}.

\begin{figure}[htb]
	\centering
	\includegraphics[scale=0.42]{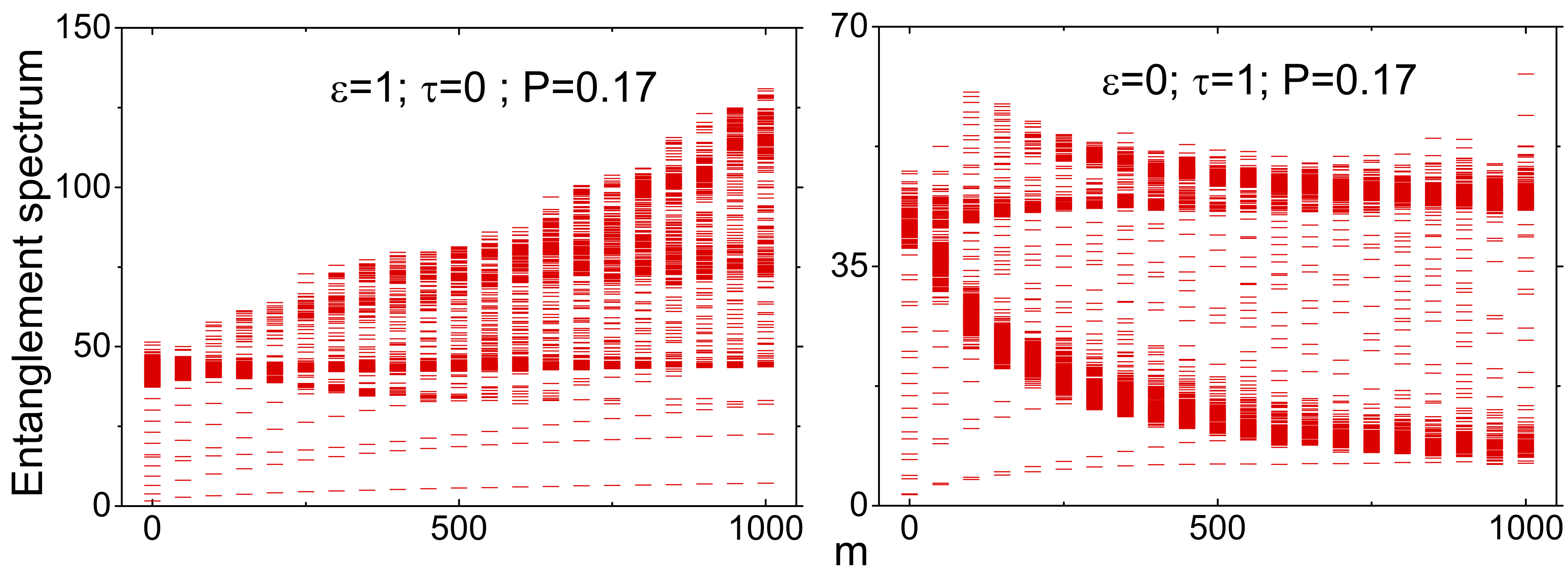}
	\caption{ Plot shows full entanglement spectrum of $h_{\rm E}$ for case I
		(left plot) and case II
		(right plot) respectively for $n=300$ and $N=600$. The red lines correspond to different entanglement energy levels. }
	\label{Fig:detailed-entangspe}
\end{figure}

Fig.~(\ref{Fig:detailed-entangspe}) contains all the eigenvalues of
the entanglement Hamiltonian $h_{\rm E}$ for the NN and NNN
coupling. It is interesting to note that in Case I, bulk of the spectrum is continuous. This has to be 
contrasted with case II where only the edges are continuous. Similar behaviour  spectrum is also seen 
in the fractional quantum hall states \cite{haldane2008} and spin chains
in momentum space \cite{Thomale2010(3)}. The fingerprint of QPTs can
be confirmed by the non-collapsing of entanglement gap --- the
entanglement energy gap between the lowest and the highest
entanglement part in the subsystem
\cite{haldane2008,Thomale2010(1),Thomale2010(2),Thomale2010(3)}.In
other words, NNN coupling term brings a gap --- that is absent in NN
coupling case --- suggesting that QPT is triggered by the linear
scalar field. This gap is generally termed as the entanglement gap or
Schmidt gap which is considered here as an order parameter to diagnose
QPT \cite{chiara2012,LAFLORENCIE2016}.
\subsection{Analytically evaluated tools}

{In this subsection we discuss the analytical results to confirm the
	evidence for violation of area-law observed in Hamiltonian
	(\ref{h2}). We use two main tools such as ground state overlap
	function or quantum fidelity and gap in the spectrum to verify the
	phase transitions in Bosonic Hamiltonian with NNN couplings.}

\subsubsection{ Ground state Overlap function and Energy gap}
The ground state overlap function (Quantum fidelity) is
defined for the ground state as $F=\langle\psi_{\rm GS}(P+\d
P)|\psi_{\rm GS}(P)\rangle$, where $\d P$ is the infinitesimal
change in the value of $P$ \cite{Anderson1967,Nielsen}.  It
has been shown that sudden change in the overlap function
indicates quantum criticality and hence is a good quantitative
measure to identify
QCP~\cite{Zanardi2006,johnpaul2008,Vieira2010}.
\begin{figure}[H]
	\centering
	\includegraphics[scale=0.6]{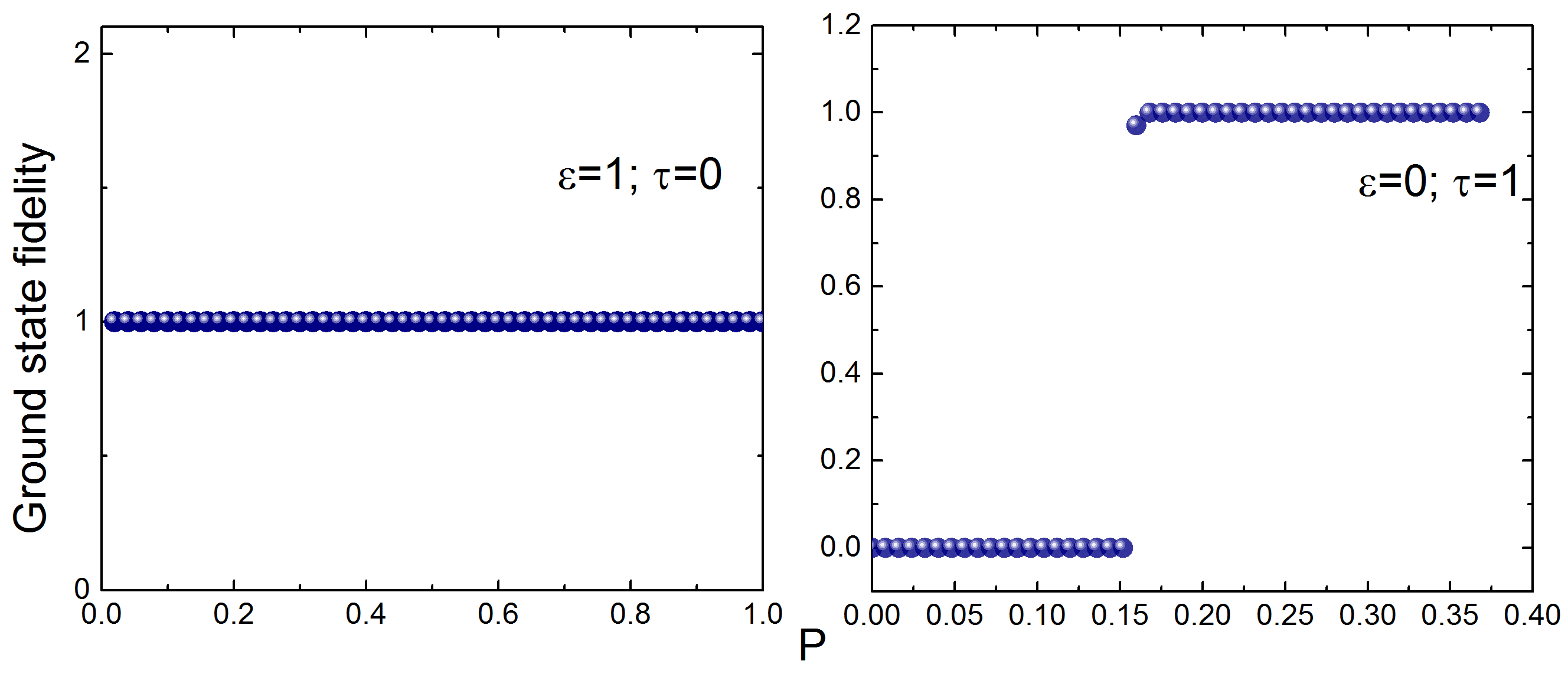}
	\caption{  Left plot is the ground state fidelity for 
		case I ($\e =1,\t=0$) while the right plot is the ground state fidelity for 
		case II ($\e =0,\t=1$). All plots are for $m=0$ and  $\d P=10^{-8}$.} 
	\label{Fig:overlap}
\end{figure}	

In Fig. (\ref{Fig:overlap}), overlap function is plotted for
different coupling parameter $P$ for the NN and NNN coupling
terms, from which we infer the following: (i) With the NN
coupling terms the overlap function is always close to unity
implying that the ground state many body wave functions for
$P$ and $P + \delta P$ are identical. This is consistent with
the results of entanglement entropy that the entropy linearly
scales with $n$ for all values of $P$.  (ii) With the NNN
coupling term, the overlap function shows a sudden
discontinuity close to $P= 0.17$. For $P > 0.17$, the ground
state wave functions (as in the case of NN coupling) for $P$
and $P + \delta P$ are identical, however, for $P < 0.17$
ground state wave functions for $P$ and $P + \delta P$ are
orthogonal.  This is also consistent with the results of
entanglement entropy that entropy is non-linearly related to
$n$. (iii) While the overlap function signals that the ground
state wave function is not identical for $P$ close of $0.17$,
entanglement entropy shows non-linear behaviour even at $P =
0.01$. This also signals that the NNN coupling term leads to
instability and leads to different phases.

It has been suggested that the signature of QPT is also
encoded in the many-body ground state of the
system~\cite{unanyan2005,Barthel2006,eisert2007}.  At quantum
criticality, ground state energy of the Hamiltonian $H(P)$ has
a gap which is related to the correlation length of the
system.  In other words, gap becomes singular --- developing
infinite correlation length across the critical point
\cite{unanyan2005,Barthel2006,eisert2007}.
\begin{figure}
	\centering
	\includegraphics[scale=.4]{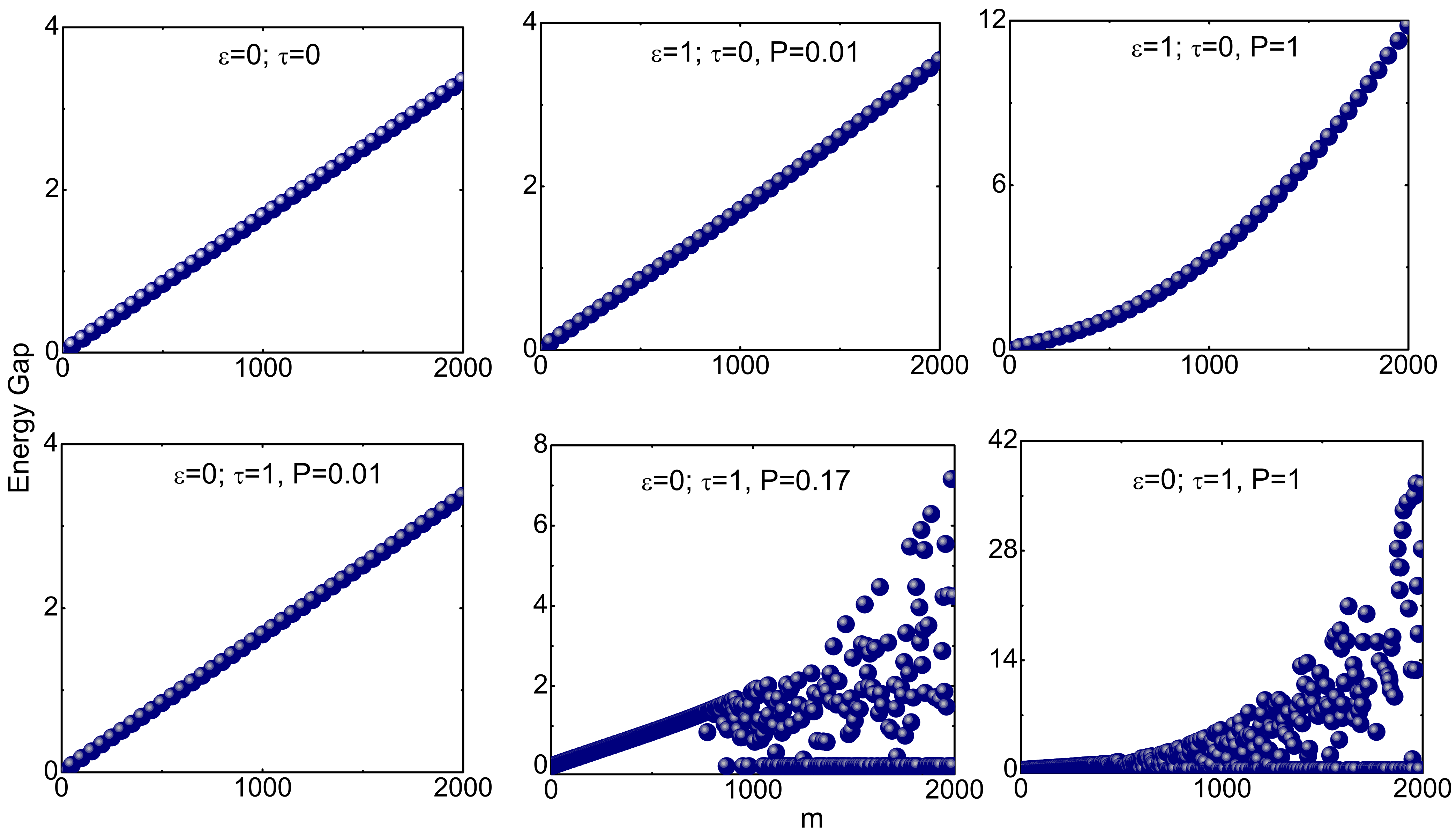}
	\caption{Plots of energy gap--- separation
		between the ground state and first excited
		state--- of $H(P)$ as a function of $m$. The
		first figure in the top panel for the
		nearest neighbouring term and the remaining
		two figures for case I ($\epsilon = 1, \tau
		= 0$) while the bottom panel is for case II
		($\epsilon = 0, \tau = 1$).}
	\label{Fig:energy}
\end{figure}	

In Fig.~(\ref{Fig:energy}), the energy gap of Hamiltonian
$H(P)$--- energy separation between the ground and first
excited states--- is plotted for different values of $m$ for
the NN and NNN coupling terms (at constant $P$), from which we
infer the following: (i) For the NN coupling term,
the energy gap vanishes only for $m = 0, P=0$.  One can infer
that only one zero mode is present, that is there is a finite
energy separation between the ground and first excited states
for any finite values of $P$.  (ii) In the case of NNN
coupling term, above the critical values of $P = 0.17$,
several values of $m$ have zero energy gap or zero mode. One
infers that multiple zero modes exist for all coupling
constants above the critical $P=(P_c=0.17)$. (iii) From the
linear behaviour of the energy versus $m$ confirms the
validity of area-law in the case of NN coupling is because of
the absence of zero mode at any non-zero $P$ values. The
accumulation of large number of zero modes at and above the
critical point $P=0.17$ characterise the violation of area-law
in the case of NNN coupling. (iv) Interestingly, the violation
of area-law can be attributed to high angular modes as one can
see it in Fig. (\ref{Fig:energy}) while after some critical
angular modes its contribution decreases, hence we can use a
cut-off while computing EE numerically. It is important to
note that this critical point is different from the point in
which the entropy-area law is violated $(P \simeq 10^{-2})$.
In other words, entanglement entropy-area violation signals a
drastic change in the ground state behaviour. Our model is yet
another example of violation of area law in higher dimensions
purely by the NNN coupling term in the Hamiltonian or the
accumulation of large number of zero modes at any non-zero $P$
values. More explicitly, the presence of multiple zero modes
lead to the violation of entropy-area relation and drives QPT
\cite{unanyan2005,Barthel2006,eisert2007}.
\section{Discussions and Future outlook}
\label{sec3}	
\subsection{What causes the quantum phase transition?}
\begin{figure} 
	\centering
	\includegraphics[scale=0.39]{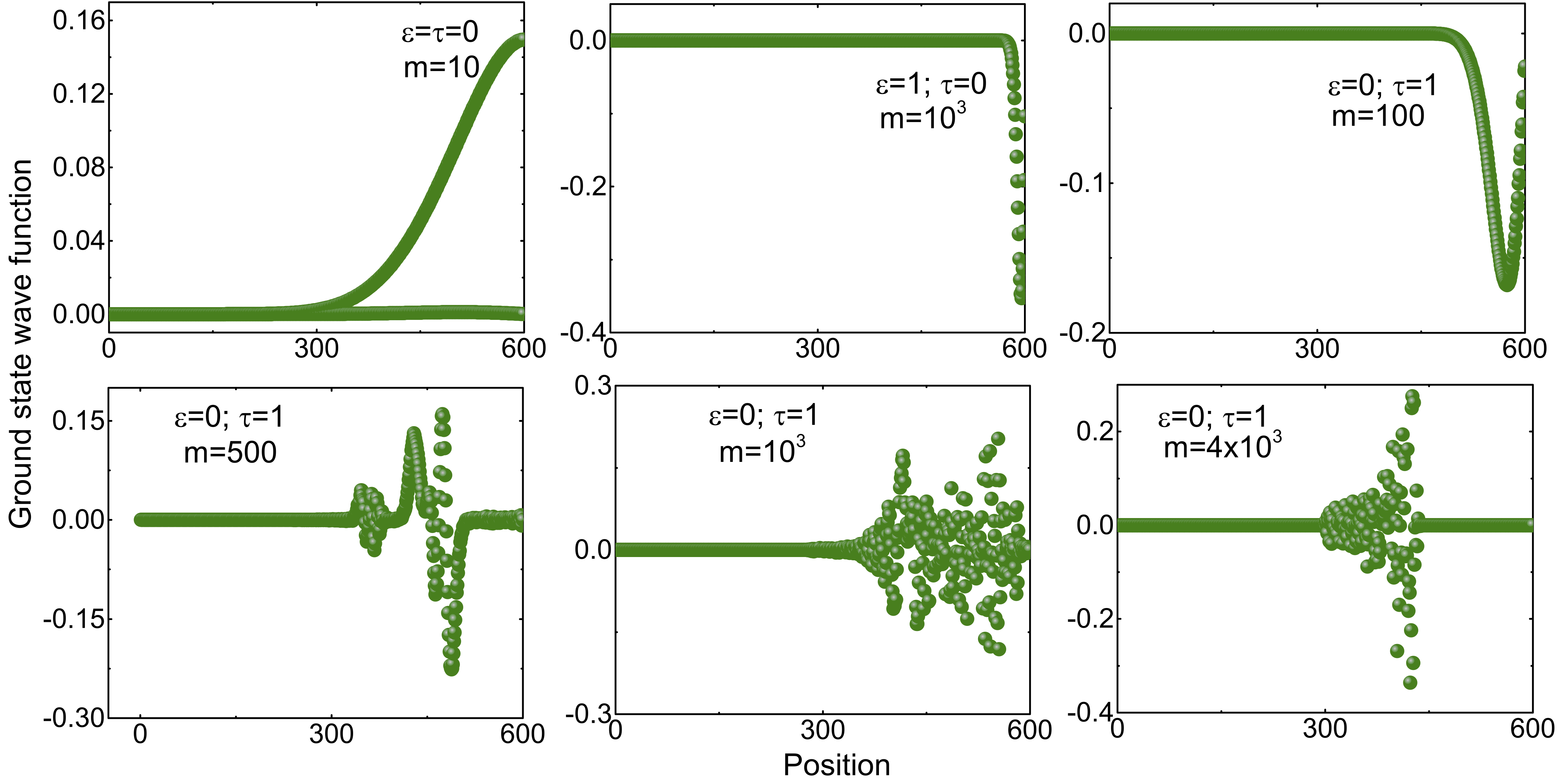}
	\caption{ Plots of ground state wave function
		versus radial distance of lattice for
		nearest neighbour (top-left), NN coupling
		term (top cent-er), and NNN coupling terms
		for $n = 300$ and $P = 1$. }
	\label{Fig:wavefunction}
\end{figure}

As we have shown explicitly, NNN coupling term leads to
QPT. It is expected that QPT should be accompanied by a
fundamental change in the ground state properties of the
system. To go about understanding this, in
Fig.~(\ref{Fig:wavefunction}), we have plotted the ground
state wave function of the system as a function of position
(lattice point). For small $m$, the wave-function peaks near
the boundary, however, for higher $m$ the wave-function is
more dispersed and peaks at the centre.

Recalling the results plotted in
Fig.~(\ref{Fig:energy}), above the critical point $P = 0.17$, 
several $m$
have zero modes and the wave-function
peaks at the centre.  The following physical
picture emerges: For small $P$, there exists
only one ($m=0$) zero mode, however, as $P$
increases the number of zero modes increase.
This constitutes a fundamental change in the
behaviour of the system and hence leading to a
phase transition. It is quite clear that for
the case of NNN coupling, the ground state
wave function is changing for each $m$ values
and contribution is coming mainly from the
bulk of the system. This change in the wave
function is responsible for the accumulation
of large number of zero modes in the system
see Fig. (\ref{Fig:energy}) for each values of
$m$. Hence, the phase transition reported in
this work is purely driven by NNN coupling and
large number of zero modes at any non-zero $P$
values implies the presence of degenerate
ground states and triggers QPT--- system
becomes gap-less
\cite{unanyan2005,Barthel2006,eisert2007}.

A qualitative understanding of excitation of higher modes (for $P> 0.17$) 
can be understood using the quantum mechanical model of a particle in a 2-dimensional box of length $L$~\cite{shanki2012}. 
It was explicitly shown that 
	\beq
	\frac{\mbox{\small Ground state energy eigenvalue for canonical scalar field}}{\mbox{\small Ground state energy eigenvalue for case II}}
	= \left[\frac{L \kappa}{\pi}\right]^{4} \label{eq:eigen-ratio}
	\eeq

Eq. (\ref{eq:eigen-ratio}) implies that, for $\kappa <\pi/L$, the
ground state energy eigenvalue for case II  is higher compared to that of canonical 
scalar field. With increasing $P$ (or decreasing $\kappa$), system readjusts in 
such a way that the ground state energy of the system increases. In other words, the cross-over
of the dispersion relation catalyses \textit{larger population} of
higher energy quantum modes.
	 
\subsection{Plausible implications in Condensed matter physics: Deriving    spatial higher  derivative Hamiltonian from Hubbard- Stratonovich transformation }

An interesting fact about the
Hamiltonian (\ref{eq:disc-Ham}) is
that it can be derived from Hubbard-
Stratonovich transformation (HST)---
Bosonization of interacting fermions
or transforming the partition
function of a system of interacting
fermions into the partition function
of noninteracting Bosons.  We briefly
sketch details of the modified
nonlinear dispersion relations from
HST for the case of $(2+1)$
dimensions \cite{HST-PRL}.

The Euclidean action, $\mathcal S^F$, for interacting electrons is,
\beq
\label{hst1}
\mathcal S^F=\int d\t_E \,d ^2{\bf x} \l(\sum_{i=\uparrow,\downarrow}\psi_i^\dagger \l[\pa_{\t_E} +\f{\nabla^2}{2 \mathcal M}\r]  \psi_i %\r.\NN\\
%&&\l.\dis
-g \psi^\dagger_\uparrow\psi^\dagger_\downarrow\psi_\downarrow\psi_\uparrow\r)
\eeq
and the partition function is 

\beq
Z= \int D[\psi^\dagger \psi] e^{-\mathcal S^F}
\eeq
where $\psi$ is the complex fermionic field,
$\t_E $ is the Euclidean time, ${\bf x}$ is
the two dimensional space vector, $ \mathcal{M}$ is the
mass of electron and $g$ is the coupling
constant. The interaction term in the above
action can be decoupled by using HST which is
given by,

\br \exp\l[-\int d \t_E \, d^2 {\bf x}\l( g
\psi^\dagger_\uparrow\psi^\dagger_\downarrow\psi_\downarrow\psi_\uparrow\r)\r]
\nn = \dis\frac{1}{\mathcal N}\int D[\Delta^*
\Delta ] \exp\l[\int d \t_E \,d^2 {\bf x}
\l(\f{1}{g} |\Delta|^2 - \Delta
\psi^\dagger_\uparrow\psi^\dagger_\downarrow
-\Delta^* \psi_\uparrow\psi_\downarrow\r)\r]
\\ = \dis\frac{1}{\mathcal N}\int D[\Delta^*
\Delta ] \exp\l[\int d \t_E \, d^2{\bf x}
\l(\f{1}{g} |\Delta- g
\psi_\uparrow\psi_\downarrow|^2 \dis- g \,
\psi^\dagger_\uparrow\psi^\dagger_\downarrow\psi_\downarrow\psi_\uparrow\r)\r]
\er
where $\mathcal N$ is the normalisation factor
that arises after the integration over
auxiliary complex field $\Delta$. The action
(\ref{hst1}) can be written in the decoupled
form after HST is, \br
\label{hst2}
\mathcal S_{HS}^F&=&\int d \t_E \,d^2 {\bf x}
\l(\sum_{i=\uparrow,\downarrow}\psi_i^\dagger
\l[\pa_{\t_E} +\f{ \nabla^2}{2 \mathcal M}\r] \psi_i
-\f{1}{g} |\Delta|^2 + \Delta
\psi^\dagger_\uparrow\psi^\dagger_\downarrow
+\Delta^* \psi_\uparrow\psi_\downarrow\r) \er
and the partition function is

\beq
Z= \int D[\psi^\dagger \psi ;\Delta^\dagger \Delta] e^{-\mathcal S_{HS}^F}
\eeq 
It should be noted that the interacting fermionic theory becomes quadratic in both $\Delta$ and $\psi$;
\br
\label{hst3}
\mathcal S_{HS}^F&=&\int d \t_E \,d ^2 {\bf x}
\l(\sum_{i=\uparrow,\downarrow}\psi_i^\dagger
\l[ G^{-1}+\Xi\r]\psi_i -\f{1}{g} |\Delta|^2+
\Delta
\psi^\dagger_\uparrow\psi^\dagger_\downarrow
+\Delta^* \psi_\uparrow\psi_\downarrow \r) \er
where we used \beq G^{-1}=
\begin{pmatrix}
	\dis 	\pa_{\t_E }  -\f{ \l(-i \nabla\r)^2}{2 \mathcal M}&0\\
	\dis 	0& \pa_{\t_E  } -\dis \f{ \l(i \nabla \r)^2}{2 \mathcal M}
\end{pmatrix}
%\eeq
\mbox{and} ~~~~~~~~
%\beq
\Xi=
\begin{pmatrix}
	0& \Delta\\
	\Delta^* &0 
\end{pmatrix},
\eeq
the space of these matrices are in the Nambu space. Performing  integration over the fermionic  fields using HST action \cite {bosonization}, 
\bea
Z&=& \int D[\psi^\dagger \psi ;\Delta^\dagger \Delta] \,e^{-\mathcal S_{HS}^F}\nn\\
&=&\f{1}{\mathcal N} \int D[\Delta^\dagger \Delta]\,Det [G^{-1}+\Xi]\,\exp\l[-\int d \t_E  \,d ^2 {\bf r} \f{1}{g}|\Delta|^2\r]\nn\\
&=&\f{1}{\mathcal N} \int D[\Delta^\dagger \Delta]  \,\exp\l[\l(\int d \t_E \, d^2 {\bf r}\, \f{1}{g}|\Delta|^2\r) + Tr \log [G^{-1}+\Xi]\ \r], 
\eea

where in the last step we have used the
matrix trace property $Det\mbox{(matrix)}=\exp (Tr\log [\mbox{matrix}])$, $Det$, and $Tr$
refer to determinant and trace operation of a matrix, respectively.

The effective action for Bosonic theory is given by;
\bea
\mathcal S_{eff}&=&\l( -\int d \t_E \, d ^2 {\bf r}  \,\f{1}{g}|\Delta|^2\r)+ Tr \log [G^{-1}+\Xi] \nn\\
&=&\l( -\int d \t_E \, d ^2 {\bf r} \,\f{1}{g}|\Delta|^2 \r)- Tr \log G^{-1} - Tr\l( G\Xi -\f{1}{2}G\,\Xi \,G\,\Xi +\ldots\r) % \r]
\eea

In the above trace expansion, all odd powers
of the combination $G\,\Xi$ are zero and the
expansion of $G^{-1} $ around the high energy
limit will bring the nonlinear dispersion
relation like $k^2 + \a k^4 + \a^2 k^6$, where
$\a$ is some dimension-full constant.  Hence,
the dispersion relation of Eq. (\ref{h2}) is
similar to the dispersion relation obtained
from the effective Bosonic Hamiltonian of some
interacting fermionic theory in the high
energy limit.  Normally, Bosonization obtained
by using HST is in the low energy regime, but
in this case it is done at the high energy
limit and we study the effects of NNN
interactions in the system.  This modified
dispersion relation is usually referred as
``trans-Bogoliubov" dispersion relation \cite
{Visser2009-prd}. Also, the Lorentz symmetry
breaking term $k^6$ plays a fundamental role
in the renormalization of the Feynman
propagators in the high energy limit (for more
details see the Ref. \cite
{Visser2009-prd}). Hence, the Hamiltonian
(\ref{h2}) plays a crucial role in the case of
strongly interacting systems and its effective
action in the high energy limit.
%(a private communication with Arturo Tagliacozzol). 

\subsection{Future outlook of the spatially coupled radial Hamiltonian}

Our analysis explicitly shows that NNN coupling term drives
QPT by modifying dispersion relation and can play an important
role in the high energy limit of the interacting Fermions. One
possible strongly correlated condensed matter system that may
be interest is high temperature superconductors (HTS). It is
long-known that, in HTS, the coloumbic interactions between
the electrons tend to make an anti-ferromagnetic arrangement
of spins in the Copper Oxide planes and the magnetic
transition is controlled by the weak coupling between the
planes along the z-axis~\cite{Bonn2006}.

To overcome the complexity of the interactions, let us
consider Bosonic scalar field $\hat{\Phi}$ in 3-dimensional
cylindrical geometry such that the higher derivative terms
contribute only in the 2-dimensional plane while the first
derivative term contribute in all the three spatial
dimensions. Repeating the analysis in 3-dimensions, it can
be shown that the model in 3-dimensions has the same entropy
profile as that in the 2-dimensional case (for detailed
calculations, see section (IV-A) of the supplementary
material).

\begin{figure}[H]
		\centering	
		\includegraphics[scale=0.5]{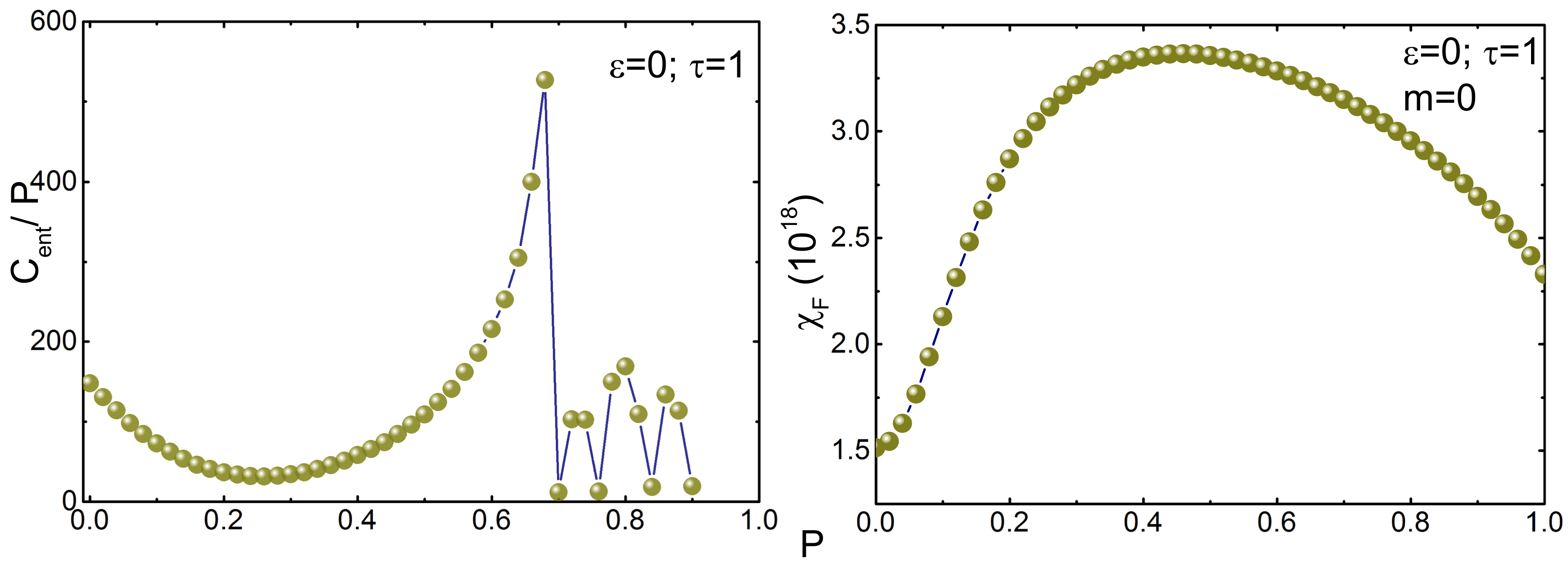}
		\caption{Left figure is the plot of entanglement specific heat
			per $P$ ($C_{ent}/P= d S_{\varsigma\rightarrow \infty}/d P
			$) versus $P$ obtained from the single copy entanglement
			entropy \cite{eisert2005} ($ S_\infty$ is the R\'{e}nyi entropy having
			infinity as the R\'{e}nyi index). Right figure is the plot
			of fidelity susceptibility ($\chi_F= \dis\lim_{\d P\to 0}-2
			\l(\d^2 /\d P^2\r) \log F$) as a function of $P$. We have taken
			lattice size of the $z$-axis to be 10 times more than the
			lattice size of the 2-dimensional surface, $N = 600, n=300$
			and $\d P=10^{-7}$. }
		\label{Fig:entangspheat}
	\end{figure}

One interesting feature is that the entanglement specific heat in Fig. (\ref{Fig:entangspheat})
	in our simplified model shows discontinuity at a particular
	value of $P$. This is indeed similar to the discontinuity of
	the specific heat measurement of the single crystals of
	YBa$_2$Cu$_3$O$_{7-\delta}$~\cite{1994-Overend.etal-PRL}.  (See
	Sec. (IVB) of the supplementary material.)
	Recently, treating entanglement as an analogue of energy, Brand\~ao
	and Plenio showed that entanglement satisfy analogous first law of thermodynamics\cite{Brandao2008}. Once a complete analogy
	between entanglement and thermodynamics is build, it may be possible
	to understand the relation between the entanglement specific heat and
	the specific heat in thermodynamics.
	While the model proposed is a simple scalar order parameter,
	it can explain some of the crucial experimental measurements
	including the discontinuity \cite{vedral2008-PRB}.  Our goal
	is to include charge carriers in the model including NNN
	coupling term and explain other phenomena in the HTS. We
	hope to report this in the future.          
		\section{ Acknowledgements}
	We thank G. Baskaran for helpful discussions and
	collaboration in the initial stage of the work. Also, it is
	our pleasure to acknowledge Samuel L. Braunstein, Jens
	Eisert, R. Ganesh, Arul Lakshminarayan, Pinaki Majumdar,
	Rajesh Narayanan, S. Pujari, G. Venketeswara Pai,  Diptiman
	Sen and Arturo Tagliacozzol for fruitful discussions and
	interesting comments. We are indeed indebted to both the
	anonymous referees for their vital comments and inputs for a
	better presentation of our work.  {We also thank Rafeeque P.P. for the help with drawing  figure (1)}. S. S. K. is grateful to
	the organizers and participants of the conference
	``Information Universe" held in Groningen, The Netherlands,
	where important progress was made.  S. S. K. is also
	thankful for the Max Planck Institute for Gravitational
	Physics, Germany and York Centre for Quantum Technologies at
	University of York, United Kingdom, for the hospitality
	during the initial and final stages of the manuscript
	respectively.  S. S. acknowledges all the participants and
	organizers of the conference ``EMN Meeting on Quantum
	Communication and Quantum Imaging", Berlin for illuminating
	discussions and comments.  All numerical codes were run in
	the high performance computer clusters at Albert Einstein
	Institute-Golm, Germany.  S. S. K. is financially supported
	by Senior Research Fellowship of the Council of Scientific
	\& Industrial Research, Government of India. This work is
	supported by Max Planck-India Partner Group on Gravity and
	Cosmology.          
	\section*{Author Contributions}
	S.S. K.  and S. S.  contributed equally for the work and   preparation of manuscript. 
	\section*{Additional Information}
	% Here, we list some of the important calculations and plots are given as an appendix of the manuscript;
%	{\bf Supplementary information} accompanies this manuscript at [url will be inserted by publisher]. \\
	%\vspace{2cm}
	\noindent{\bf Competing financial interests: }The authors declare no competing financial interests.  
		\newpage
\bc {\bf \LARGE---Supplementary Material---}\ec	
\setcounter{section}{0}	 
	       \section{  The interaction matrix elements in two dimensional space}
	\label{app1}
	The  $2-$ dimensional higher spatial derivative Hamiltonian is \cite {shanki2012}
	
	\beq
	\label{h1}
	\hat H=\frac{1}{2}\int{d^2 {\bf r}  \left[|\hat\Pi({\bf r})|^2 +\left| \nabla \hat\Phi({\bf r})\right| ^2 +\frac{\e}{\k^2}\l|\nabla ^2\hat\Phi({\bf r})\r|^2
		+\frac{\t}{\k^4}\l|\nabla ^3\hat\Phi({\bf r})\r|^2\right]} 
	\eeq
	
	\noindent where  $\hat\Phi$  is the massless scalar field  and $\hat\P$ is its conjugate momenta, $\epsilon$ and $\tau$  are  dimensionless constants, 
	$\kappa$ has the dimension of wave number and ${\bf r} (=r,\th) $ is the circular polar coordinates.  All  computations are  done   by setting $\hbar =c=1$.
	
	The equal time canonical  commutation relation is given by,
	\beq
	\left[\hat\Phi({\bf r}),\hat\Pi({\bf r'} )\right] = i\,\delta^2 ({\bf r-r'})= \frac{i}{r}\delta ( r-r')\delta ({ \th-\th'})
	\eeq
	
	We use the following ansatz to expand the real scalar fields in circular polar coordinates 
	
	\begin{subequations}
		
		\br
		\hat\Pi( {\bf r})= \sum_{m=-\infty}^\infty \frac{\hat\Pi_m(r)}{\sqrt{\pi r}}\cos m \theta \\
		\hat\Phi({\bf r})= \sum_{m=-\infty}^\infty \frac{\hat\vph_m(r)}{\sqrt{\pi r}}\cos m \theta,
		\label{h}
		\er
	\end{subequations}	
	
	where $m$ is the angular momentum quantum number.
	The canonical commutation relation between the  new rescaled fields is 
	%fields is  between the components of the Hermitian fields in Eq:(\ref{h}) is
	\beq
	\left[\hat \vph_m({ r}),\hat \Pi_{m'}({ r'} )\right] = i \, \delta ({ r-r'})\thinspace \delta_{mm'}
	\eeq
	
	Integration over the  polar angle $\theta$ is carried out by invoking the orthogonal properties of the 
	cosine function. We then apply the central difference scheme on a radial lattice having $N$ lattice points with $a$ is the lattice parameter.
	The Hamiltonian in eq. (\ref{h1}) becomes, 
	
	%	The Hamiltonian in Eq. (\ref{hm}) is discretized as a set of N coupled harmonic oscillators via the central difference scheme (CDS)  and is given by
	%with $\displaystyle H=\sum_{m} H_m $ where $H_m$ 
	\br
	\label{hmm}
	\hat H &= &\frac{1}{2 a}\sum_{m=-\infty}^{\infty}\sum_{j=1}^N\left[\hat\pi^2 _{m,j} +  \l(\frac{\hat\vph_{m,j+1} -\hat\vph_{m,j-1}}{2}\right)^2  -\frac{\hat\vph_{m,j}}{2 j} \left(\hat\vph_{m,j+1} -\hat\vph_{m,j-1}\right) 
	+  (1+4 m^2) \frac{\hat\vph^2_{m,j}}{ j^2} \r.\nn\\
	&&	\l. +\frac{\e}{P}\l\{\l(\hat\vph_{m,j+1}-2\hat\vph_{m,j}+\hat\vph_{m,j-1}\r)^2
	+\frac{2 \b}{j^2}\l(\hat\vph_{m,j+1}-2\hat\vph_{m,j}+\hat\vph_{m,j-1}\r)^2
	+\frac{\b^2}{j^4}\hat\vph_{m,j}^2\right\}\r.\nn\\
	&&\l. +\frac{\t}{P^2}\l\{\l(\frac{\hat\vph_{m,j+2}-2\hat\vph_{m,j+1}+2\hat\vph_{m,j-1}-\hat\vph_{m,j-2}}{2}\r)^2+\frac{\a}{j^2}\l(\hat\vph_{m,j+1}-2\hat\vph_{m,j}+\hat\vph_{m,j-1}\r)^2 \r.\r.\nn\\
	&&\l.\l. +\frac{\b^2}{j^4}\l(\frac{\hat\vph_{m,j+1}-\hat\vph_{m,j-1}}{2}\r)^2 
	+\frac{25/4+m^2}{j^6}\b^2\hat\vph_{m,j}^2-\frac{5\b^2 }{2j^5}\l(\frac{\hat\vph_{m,j+1}-\hat\vph_{m,j-1}}{2}\r)\hat\vph_{m,j}\r.\r.\nn\\
	&&\l.\l.-\l(\frac{\hat\vph_{m,j+2}-2\hat\vph_{m,j+1}+2\hat\vph_{m,j-1}-\hat\vph_{m,j-2}}{2j}\r)\l(\hat\vph_{m,j+1}-2\hat\vph_{m,j}+\hat\vph_{m,j-1}\r)\r.\r.\nn\\
	&&\l.\l.+ \frac{2\b}{j^2}\l(\frac{\hat\vph_{m,j+2}-2\hat\vph_{m,j+1}+2\hat\vph_{m,j-1}-\hat\vph_{m,j-2}}{2}\r)\l(\frac{\hat\vph_{m,j+1}-\hat\vph_{m,j-1}}{2}\r)\r.\r.\nn\\
	&&\l.\l.- \frac{5\b}{j^3}\l(\frac{\hat\vph_{m,j+2}-2\hat\vph_{m,j+1}+2\hat\vph_{m,j-1}-\hat\vph_{m,j-2}}{2}\r)\hat\vph_{m,j}-\frac{\b}{j^3}\l(\hat\vph_{m,j+1}-2\hat\vph_{m,j}+\hat\vph_{m,j-1}\r)\r.\r.\nn\\
	&&\l.\l.\times\l(\frac{\hat\vph_{m,j+1}-\hat\vph_{m,j-1}}{2}\r)
	+\frac{\l(5/2+2m^2\r)\b}{j^4}\l(\hat\vph_{m,j+1}-2\hat\vph_{m,j}+\hat\vph_{m,j-1}\r)\hat\vph_{m,j} \r\}\r]
	\er
	where  $\pi_{m,j}=a\;\Pi_{m,j}$ is the new canonical momentum field,  and $\hat\vph_{m,N+1}=0$ is the constraint on field.
	
	%and $ \hat\Phi_{m,j},\hat\Pi_{m,j}$ are the dimensionless field satisfies
	The  canonical commutation relation between the dimensionless fields is 
	\beq
	\left[\hat \vph_{m,j},  \hat\pi_{m',j'} \right] = i \delta_{mm'} \delta_{jj'}
	\eeq 
	%\end{subequations}

	More explicitly we can write the above interacting quantum  Hamiltonian as  a set of $N$ coupled harmonic oscillators with time independent frequency as \cite{bombelli86,srednicki93}, 
	\beq
	\label{hk}
	H(P)=\frac{1}{2a}\sum_{m=-\infty}^{\infty} \sum_{i,j=1}^N \l[\pi_{m,i}^2\;\d_{i,j} +  \phi_{m,i}\; K_{ij}(P,m) \;\phi_{m,j}\r]
	\eeq  
	where $K_{ij}$  is a real symmetric interaction  matrix have positive real energy eigenvalues and is given by,
	
	\br
	K_{ij}(P,m)& =& K^{(1)}\delta_{1,j}+K^{(2)}\delta_{2,j}+K^{(3)}\delta_{i,j (j\ne 1,2,N-1,N)}+K^{(4)}\delta_{N-1,j}+K^{(5)}\delta_{N,j}\nn\\
	&&\dis K^{(6)}\l(\delta_{1,j-1}+\delta_{2,j+1}\r)+\dis K^{(7)}\l(\delta_{i,j-1(j\ne2)}+\delta_{i,j+1 (i\ne 2)}\r)+\dis K^{(8)}\l(\delta_{i+1,N}+\delta_{N,j+1}\r) \nn\\
	&&+ K^{(9)}\l(\delta_{i+2,j}+\delta_{i,j+2}\r)+K^{(10)}\l(\delta_{i+3,j}+\delta_{i,j+3}\r)+K^{(11)}\l(\delta_{i+4,j}+\delta_{i,j+4}\r)
	\er                               
	where 
	\begin{subequations}                        
		\br
		K^{(1)}&=&\dis\frac{1}{4}+\frac{\alpha}{j^2} + \epsilon P \left(5-\frac{4 \beta}{j^2}+ \frac{\beta^2}{j^4}\right) +\tau P^2 \left(\frac{3}{4}+\frac{17 \alpha}{(j+1)^2} + 
		\frac{\gamma \beta^2}{j^6}-\frac{2 \varsigma \beta}{j^4}-\frac{3 \beta}{2(j+1)^3}   +\frac{\beta^2}{4(j+1)^4}\right)\\
		K^{(2)}&=& \dis\frac{1}{2}+\frac{\alpha}{j^2} + \epsilon P \left(6-\frac{4 \beta}{j^2}+ \frac{\beta^2}{j^4}\right)+\tau P^2 \left(\frac{35}{12}                                  +\frac{19\alpha}{(j+1)^2}+ \frac{\gamma \beta^2}{j^6} -\frac{2 \varsigma \beta}{j^4}-\frac{43 \beta}{(j+1)^3}+\frac{82\beta^2}{4(j+1)^4}\right)   \\   
		K^{(3)}&= & \dis\frac{1}{2}+\frac{\alpha}{j^2}+ \epsilon P \left(6-\frac{4 \beta}{j^2}+ \frac{\beta^2}{j^4}\right)+\tau P^2 \left(\frac{5}{2} +\frac{2}{j^2-1}+\frac{4 \alpha}{j^2} +\frac{\gamma \beta^2}{j^6}-\frac{2 \varsigma \beta}{j^4}+\left( \alpha-\beta \right)\right.\nn\\
		&&\left.\dis \times\left(\frac{1}{(j+1)^2}+\frac{1}{ (j-1)^2}\right)+\frac{\beta^2}{4} \left(\frac{2}{(j+1)^4}   +\frac{1}{ (j-1)^4}\right)+\frac{\beta}{2} \left(\frac{1}{(j+1)^3}-\frac{1}{(j-1)^3}\right)\right)\\     
		K^{(4)}&= &    \dis\frac{1}{2}+\frac{\alpha}{(N-1)^2}+ \epsilon P \left(6-\frac{4 \beta}{(N-1)^2}+ \frac{\beta^2}{(N-1)^4}\right)+\tau P^2 \left(\frac{9}{4}+\frac{2}{(N-2)N} +\left( \alpha-\beta \right) \right.\nonumber\\
		&& \left.\dis \times \left(\frac{1}{(N-2)^2}+\frac{1}{ N^2}\right)+\frac{\beta^2}{4} \left(\frac{1}{(N-2)^4}+\frac{2}{N^4}\right)+\frac{4 \alpha}{(N-1)^2} +\frac{\gamma \beta^2}{(N-1)^6} -\frac{2 \varsigma \beta}{(N-1)^4}\right. \nonumber\\  
		&&\left. \dis+\frac{\beta}{2} \left(\frac{1}{N^3}-\frac{1}{(N-1)^3}\right)\right)   \\
		K^{(5)}&= &  \dis \left(\frac{1}{4}+\frac{\alpha}{N^2}\right)+\epsilon P \left(5-\frac{4 \beta}{N^2}+ \frac{\beta^2}{N^4}\right) +\tau P^2 \left(\frac{5N+1}{4(N-2)}+\frac{\alpha-\beta}{(N-1)^2}+\frac{\beta^2}{4(N-1)^4}-\frac{\beta}{2(N-1)^3} \r.\nn\\
		&&\l.\dis
		+\frac{\gamma \beta^2}{N^6}-\frac{2 \beta\varsigma}{N^4}\right)  \\
		K^{(6)}&= &  \dis    -\frac{1}{8}+\epsilon P\left(\frac{5\beta}{4}-4\right)+\frac{\tau P^2}{2}\left(\frac{-11}{6}-5\alpha+\frac{191\beta}{36}+
		\frac{17\beta\varsigma}{16} -\frac{155\beta^2}{64}\right)   \\
		K^{(7)}&= &  \dis -\frac{1}{4i(i+1)}+\epsilon P\left(-4+\beta\left(\frac{1}{i^2}+\frac{1}{(i+1)^2}\right)\right)+\frac{\tau P^2}{2}\left(-2-\frac{3}{(i-1)(i+2)} \right.\nonumber\\
		&&\left.-\frac{1}{i(i+1)}+\frac{\beta}{2(i-1)^2}-\frac{4\alpha}{i^2}+\frac{\varsigma\beta}{i^4}+\frac{6\beta}{i^3}-\frac{5\beta^2}{2i^5}-\frac{4\alpha}{(i+1)^2}+\frac{\varsigma\beta}{(i+1)^4}+\frac{6\beta}{(i+1)^3} \right.\nonumber\\
		&&\left. -\frac{5\beta^2}{2(i+1)^5}-\frac{4\alpha}{(i+1)^2}+\frac{\beta}{2(i+1)^2}\right)\\
		K^{(8)}&= &  K^{(7)}+\left(1-\frac{1}{2(N+1)}-\frac{\beta}{2(N+1)^2}\right)\\
		K^{(9)}&= &  -\frac{1}{4}+\epsilon P +\frac{\tau P^2}{2}\left(-2+\frac{2\alpha}{(i+1)^2}-\frac{\beta^2}{2(i+1)^4}+\frac{2}{i(i+2)}+\frac{5\beta}{2}\left(\frac{1}{(i+2)^3}-\frac{1}{i^3}+\frac{4\beta}{5(i+1)^2}\right)\right)\\
		K^{(10)}&= & \frac{\tau P^2}{2}\left[2-\frac{1}{(i+1)(i+2)}-\frac{\beta}{2}\left( \frac{1}{(i+1)^2}+\frac{1}{(i+2)^2}\right)\right]  \\
		K^{(11)}&= & -\frac{\tau P^2}{4}                                                       
		\er   
	\end{subequations}                 
	and                     \begin{subequations}
		\br
		\alpha=\frac{1+4 m^2}{4},~~
		\beta=\frac{1-4m^2}{4},~~
		\gamma=\frac{25}{4}+m^2,~~
		\varsigma=\frac{5}{2}+2 m^2\er 	
	\end{subequations}   
	It is important to note that the interaction matrix $K_{ij}$ contains all the diagonal entries ($K^{(1)},K^{(2)},K^{(3)},K^{(4)},K^{(5)}$), 
	nearest neighbour couplings ($K^{(6)},K^{(7)},K^{(8)}$), NN couplings ($K^{(9)}$) and NNN couplings ($K^{(10)},K^{(11)}$). The NNN 
	couplings are from the sixth order derivative term.                                         
	\begin{figure}
		\centering
		\includegraphics[scale=.35]{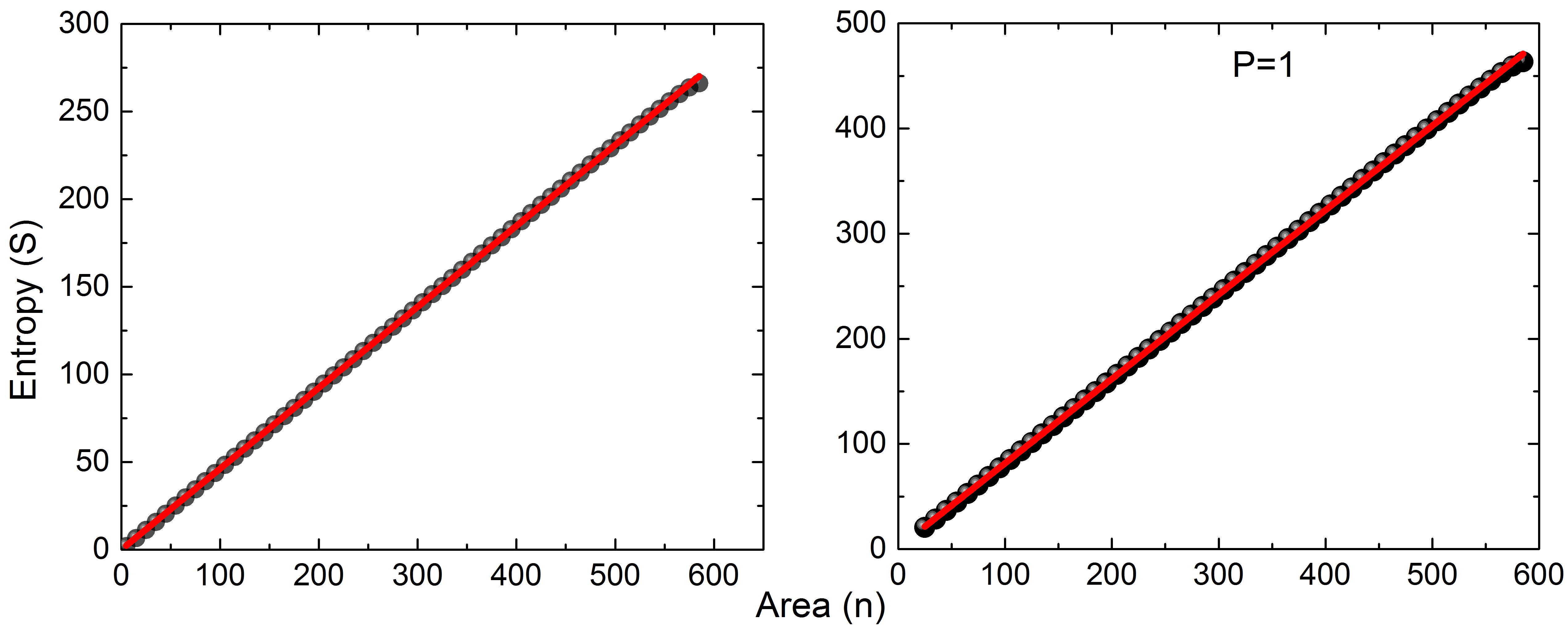}
		\caption{Left figure is the plot of  entanglement entropy $(S)$  vs $n$ for linear dispersion relation, $ \omega =k$. Right
			figure is the plot of entanglement entropy $(S)$  vs $n$ for $\omega^2  = k^2 + k^4/\kappa^2$. In all cases we set $N=600$.  
			The black dots represent numerical data points and the red line is the best linear fit represented 
			by the equation $S=a\thinspace n\thinspace+0.038\thinspace\log n $, where $a$ is slope of the line for both the cases. 
			This is consistent with the results of \cite{fradkin2006prl}. The expression for the entropy, the fitted red line, 
			is a universal property of entanglement entropy for $z=2$ class systems.}
		\label{Supp:fig1}
	\end{figure}
	
	We evaluate the entropy using the method discussed in Refs. \cite{bombelli86,srednicki93} using the von-Neumann expression as $S=-\mbox{Tr} \,\rho \log \rho$. In all numerical computations are done in  MATLAB R2012a,
	we use $10^{-8}$ as a numerical accuracy   and all  matrix entries in the $K_{_{ij}}$ are rescaled by a factor of $10^{-10}$.  Figure (\ref{Supp:fig1}) 
	contains the plot of EE versus $n$ for the linear dispersion relation ($\omega = k$) (left plot) and $\omega^2  = k^2 + k^4/\kappa^2$ (right plot).  
	
	It is interesting to note that EE is proportional to area  of the boundary, that is $n$  for the case of nearest neighbour and NN coupling, 
	however, its profile completely changes in the case of NNN coupling. 
	%             \begin{figure}
	%                      	\centering
	%                      	\includegraphics[scale=.38]{entropyp1}
	%                      	\caption{This is the  plot of  entanglement entropy $(S)$  vs $n$ only for quadratic dispersion relation, $ \omega = k^2/\k$ and $ N=600$. 
	%                      		The black dots represent numerical data points and the red line is the best linear fit represented 
	%                      		by the equation $S=1.39\thinspace n\thinspace+0.038\thinspace\log n $. This is exactly consistent with the analysis of E. Fradkin and J. E. Moore reported in Ref. \cite{fradkin2006prl}. The expression for the entropy, the fitted red line, is a universal property of entanglement entropy for $z=2$ class systems. }
	%                      	\label{Supp:fig4}
	%                      \end{figure}
	%  \newpage     
	\section{ Entanglement entropy for different lattice sizes}
	\label{app2}
\begin{figure}
		\centering
		\subfigure[]{%
			\includegraphics[scale=.33]{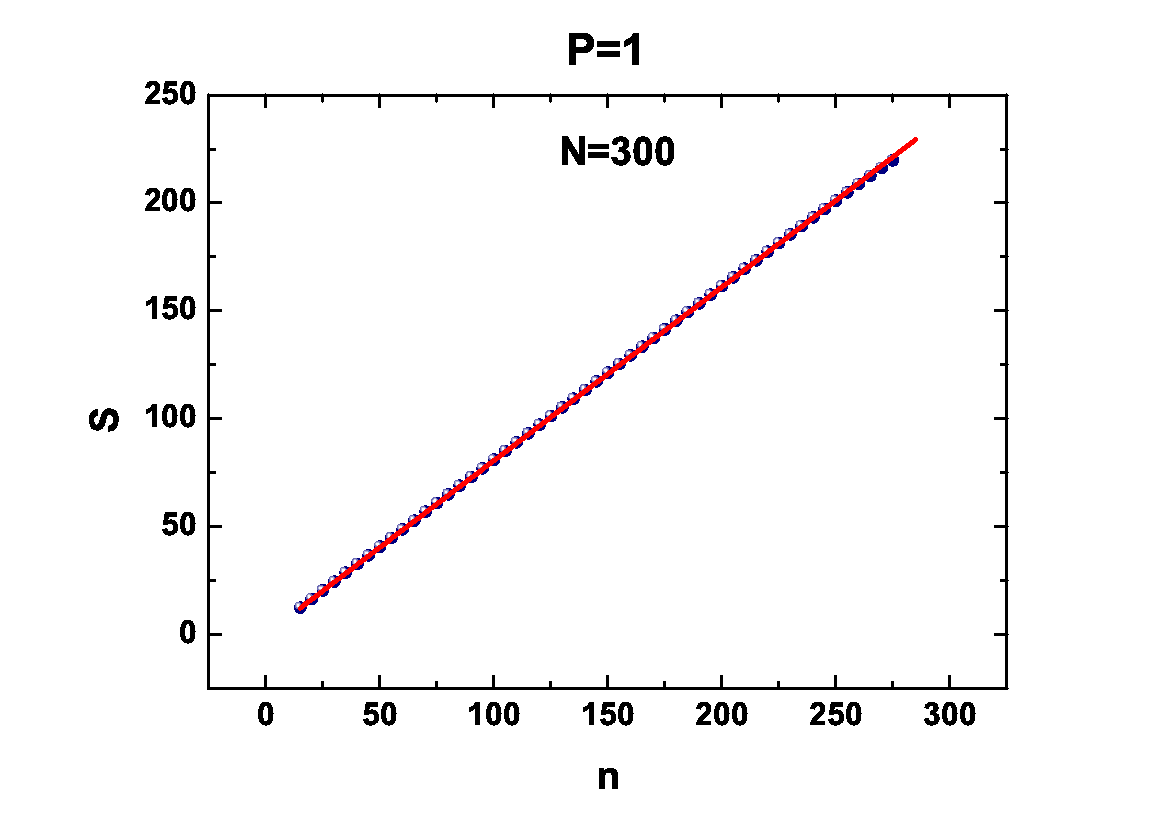}
		}%
		%		\subfigure[]{%
		%			\includegraphics[scale=.3]{fig24}
		%		}\\ %? ------- End of the first row ----------------------%
		\subfigure[]{%
			\includegraphics[scale=.33]{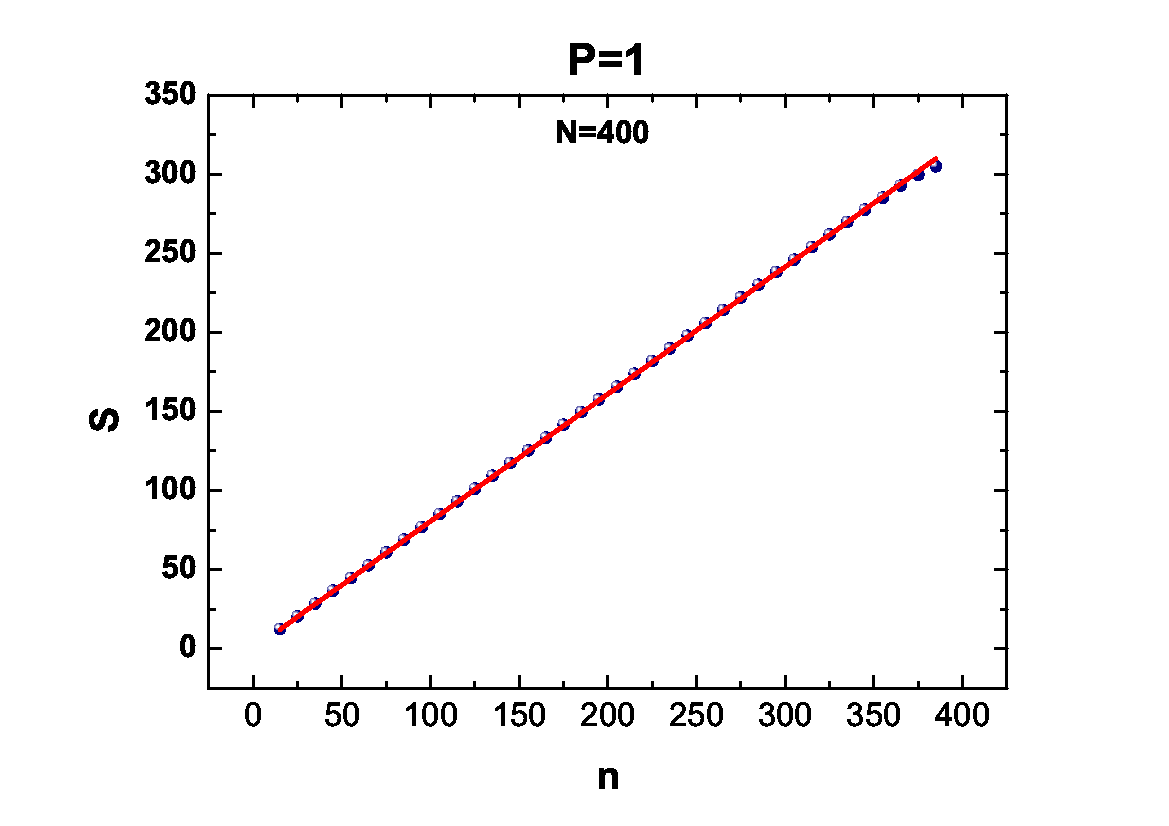}
		}\\%
		\subfigure[]{%
			\includegraphics[scale=.33]{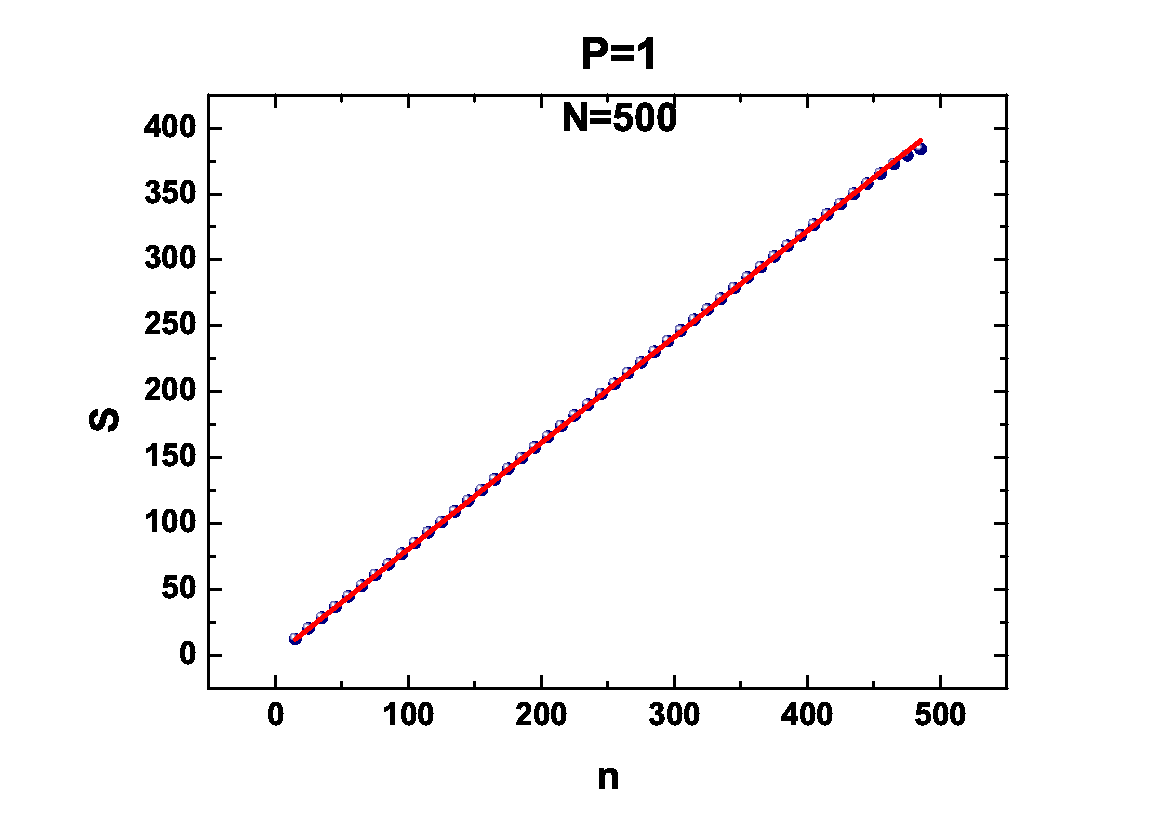}
			%\subcaption{hh}
		}%
		\caption{ Plot of entropy as a function of n for various $P$ values for $\epsilon=1,\tau=0$ with N=300,400, and 500  sites respectively arranged in 
			sub figures (a), (b), and (c). The blue dots in each plot represents numerical data and the red line is  the best linear fit represented by 
			$S=a\thinspace n\thinspace+0.038\thinspace\log n $, where $a$ is slope of the line. These plots are consistent with the analysis reported in Ref. \cite{fradkin2006prl}}
		\label{Supp:fig2}
	\end{figure}

\begin{figure}
		\centering
		\subfigure[]{%
			\includegraphics[scale=.3]{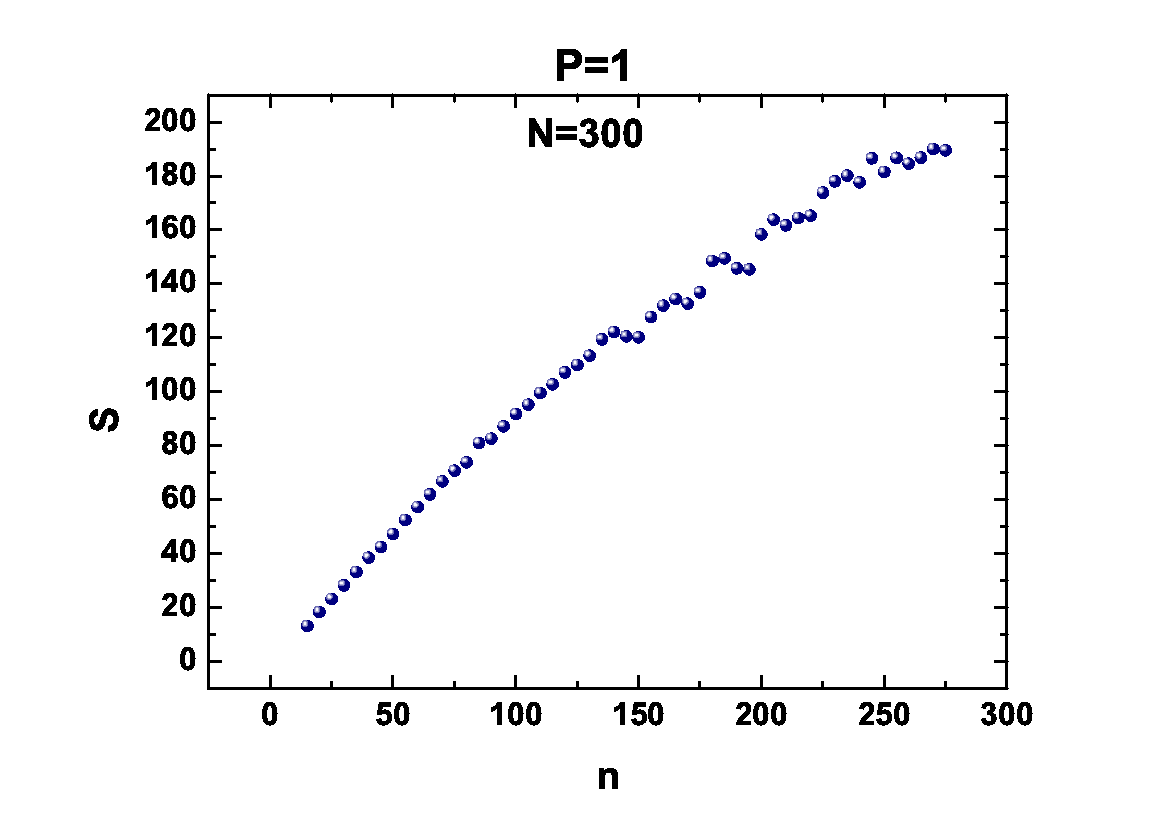}
		}%
		\subfigure[]{%
			
			\includegraphics[scale=.3]{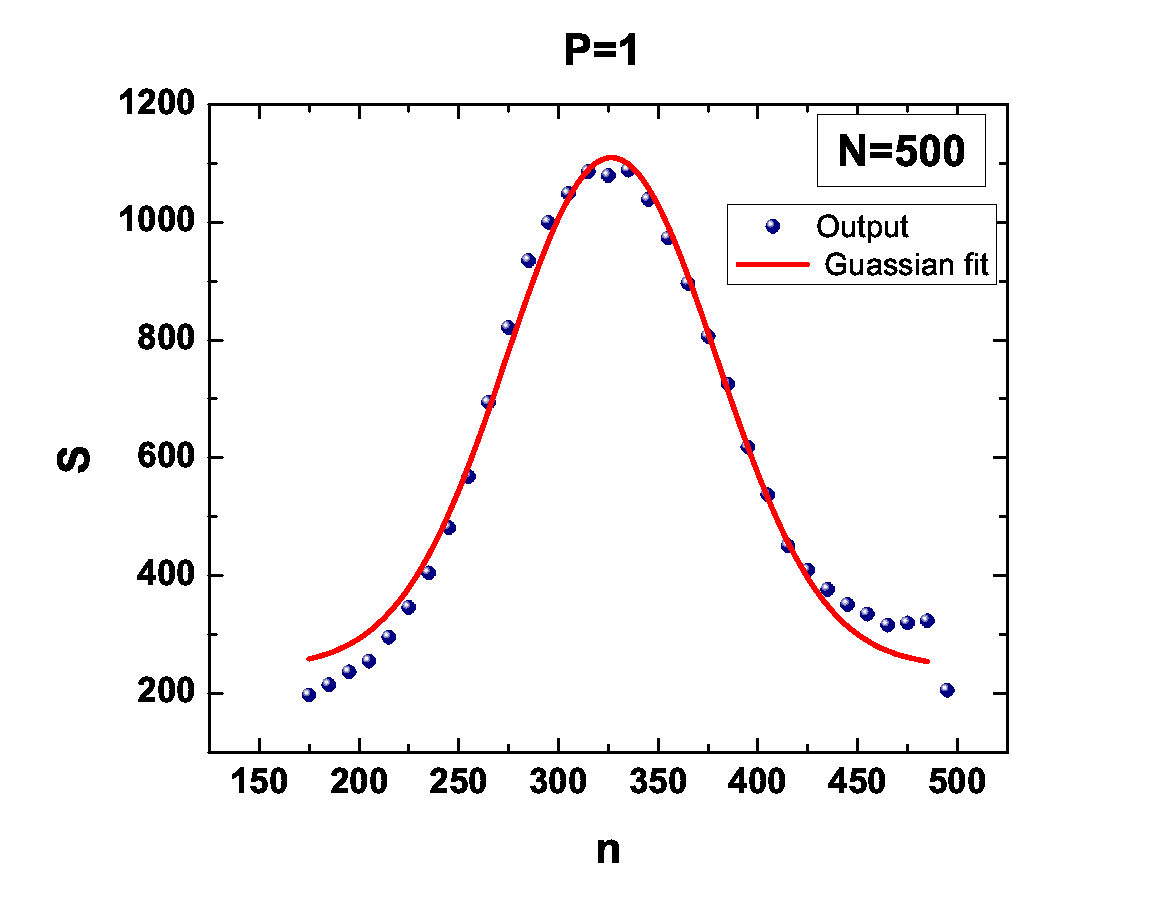}
		} %? ------- End of the first row ----------------------%
		\subfigure[]{%
			\includegraphics[scale=.3]{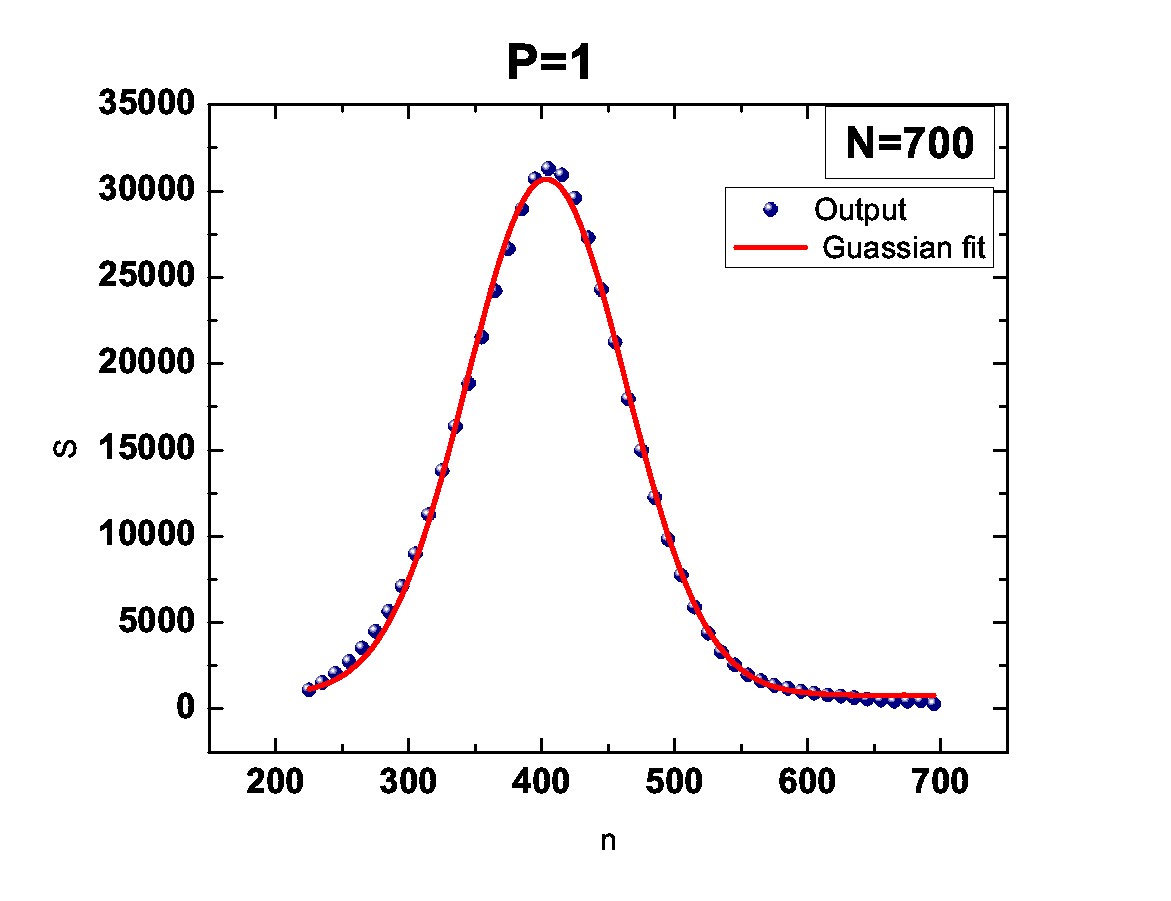}
		}%
		
		\caption{ Plot of entropy as a function of $n$ for various $P$ values for $\epsilon=1,\tau=1$ with $N=300, 500, \mbox{and}  ~700$ sites respectively arranged in sub figures (a), (b), and (c). The blue dots 
			represent the output and  red colour is  the best fit in figs. (b) and (c).}
		\label{Supp:fig3}
	\end{figure}

	The following figures represent the behavior of EE for different $N$'s.
	Plots in Fig. (\ref{Supp:fig2}) are for the dispersion relation $ \omega^2  = k^2 + k^4/\kappa^2$ and the plots in Fig. (\ref{Supp:fig3}) are 
	for the dispersion relation $ \omega^2  = k^2 + k^4/\kappa^2 + k^6/\kappa^4$. It is clear for the former case that the entropy scales with 
	area and the next leading order is the  diverging logarithmic correction which was reported for $z=2$ critical theories in two dimensions \cite{fradkin2006prl}. 
	However, in case (II), the we observe a violation in area-law due to the NNN couplings and it is shown in Fig. (\ref{Supp:fig3}) for various $N$ values. 
	
	\newpage
	\section{Entanglement entropy of a cylinder}
	The  Hamiltonian in the 3-dimensional cylindrical geometry is given by:
	\beq
	\label{h1-cylinder}
	\hat H_1=\frac{1}{2}\int{d^2 {\bf r} \;dz \left[\hat\Pi^2({\bf r},z) +\left| \nabla \hat\Phi({\bf r},z)\right| ^2 +\frac{\e}{\k^2}\l|\nabla ^2_{{\bf r}}\hat\Phi({\bf r},z)\r|^2
		+\frac{\t}{\k^4}\l|\nabla ^3_{{\bf r}}\hat\Phi({\bf r},z)\r|^2\right]} 
	\eeq
	\noindent where  $\hat\Phi$  and $\hat\P$ are the one component  real massless scalar field and its conjugate momenta respectively, 
	$\epsilon$ and $\tau$  are  dimensionless constants, $\kappa$ has the dimension of wave number and ${\bf r} (=r,\th) $ is the circular polar coordinates and 
	$z$ is the height of the cylinder. Here $\nabla ^2_{{\bf r}} $ and $\nabla ^3_{{\bf r}}$ represent  higher order derivative operators for the 2-D plane. That is 
	we study the effect of the higher derivative terms  in the 2-D plane  ---- interaction strength is comparatively less in the $z$ direction. 
	
	The equal time canonical  commutation relation between fields, which are the functions of $\bf r$ and $z$  is given by,
	\beq
	\left[\hat\Phi({\bf r},z),\hat\Pi({\bf r'},z )\right] = i\,\delta^2 ({\bf r-r'})\;\delta ( z-z')= \frac{i}{r}\delta ( r-r')\;\delta ({ \th-\th'})\;\delta ( z-z')
	\eeq
	
	We use the following ansatz for expanding the real scalar fields in circular cylindrical coordinates 
	
	\begin{subequations}
		
		\br
		\hat\Pi( {\bf r},z)=\frac{1}{2\pi^2}\int_0^\infty dk \sum_{m=-\infty}^\infty \frac{\hat\Pi_m(r,k)}{\sqrt{ r}}\cos m \theta \;\cos k z\\
		\hat\Phi({\bf r},z)= \frac{1}{2\pi^2}\int_0^\infty dk \sum_{m=-\infty}^\infty \frac{\hat\vph_m(r,k)}{\sqrt{ r}}\cos m \theta\;\cos k z,
		\label{h-cylinder}
		\er
	\end{subequations}	
	
	where $m$ is the angular momentum quantum number and $k$ is a real positive parameter which has the dimension of momentum.
	The canonical commutation relation between the  new rescaled fields is 
	%fields is  between the components of the Hermitian fields in Eq:(\ref{h}) is
	\beq
	\left[\hat \vph_m({ r,k}),\hat \Pi_{m'}({ r',k'} )\right] = i \, \delta ({ r-r'})\, \delta  (k-k')\thinspace \delta_{mm'}
	\eeq
	
\begin{figure}
		\begin{center}
			\subfigure[]{%
				\includegraphics[scale=.35]{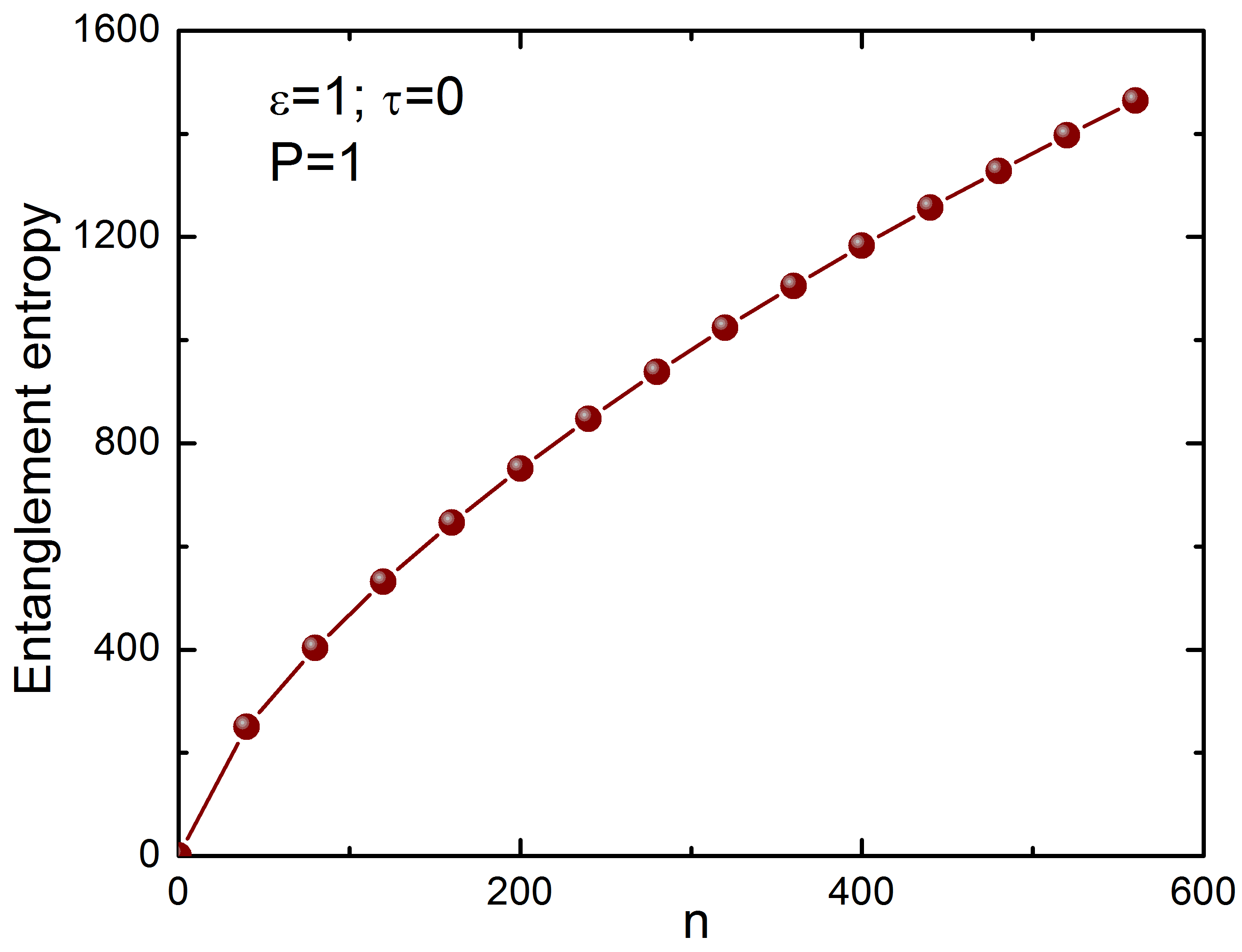}
			}%
			\subfigure[]{%
				\includegraphics[scale=.35]{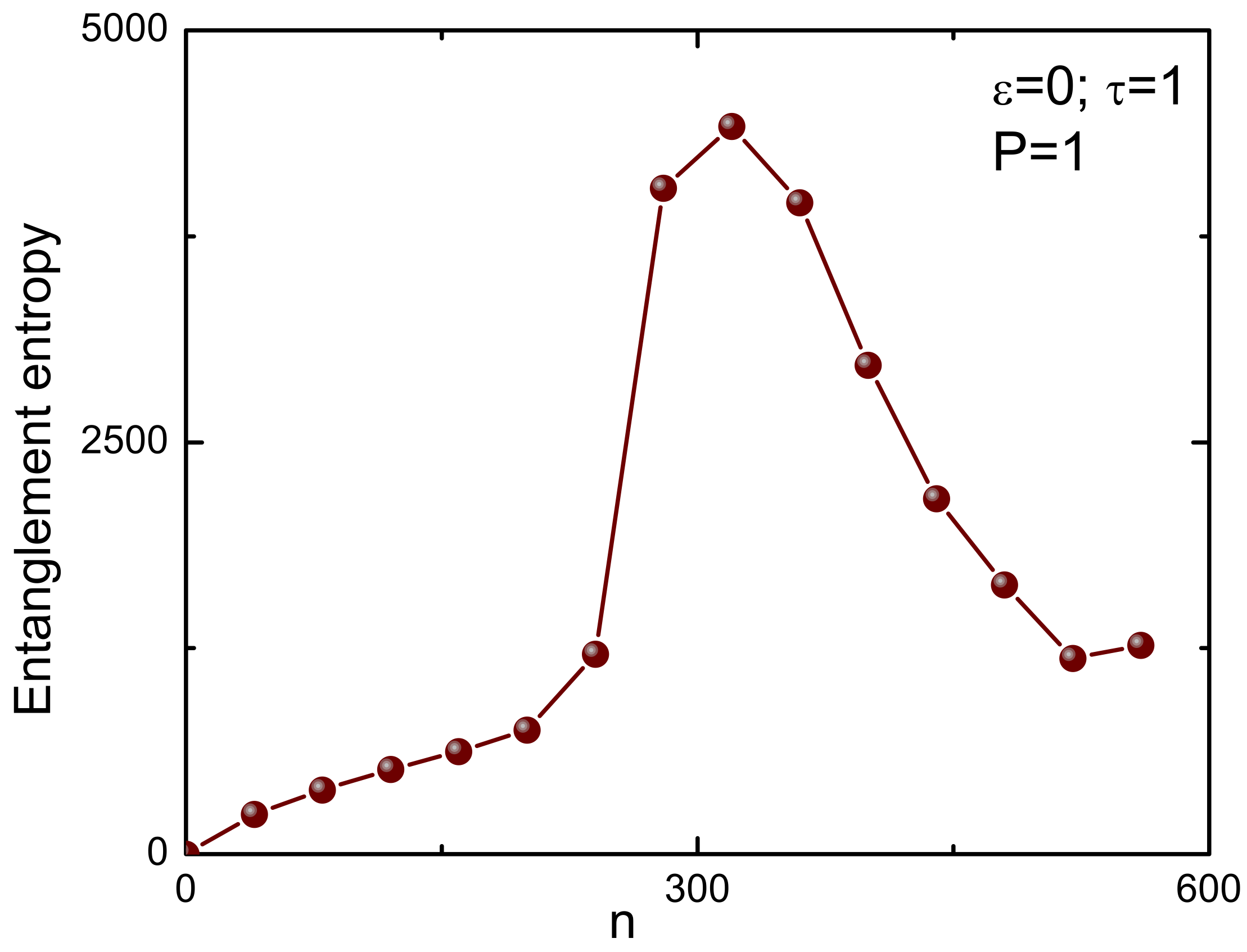}
			} %? ------- End of the first row ----------------------%
		\end{center}
		\caption{ Left figure is the plot of entanglement entropy versus area for scenario I.  Right figure is the plot of entanglement entropy versus area for scenario II.  
			We have used the lattice size in the z-axis is 10 times the lattice size in the 2-D plane. The number of lattice points used in the calculation are $N=600,N_1=2, \d P=10^{-9}$ and 
			the accuracy is $10^{-2}$.  }
		\label{Supp:cylinderentropy}
	\end{figure}

	Unlike in the case of circular polar coordinates, here we have to do two different discretization \textemdash one for the radial direction and other along in the height of the cylinder. For numerical simplicity, we choose the height of the cylinder is some arbitrary number $\nu$ times its radius. Let $a$ and $b$ are the lattice parameter in the radial and $z$- direction respectively and here we takes $b= \nu\; a$. That is,
	\beq
	r=j a ;\qquad\qquad  1/k=s \;\nu\; a
	\eeq
	After some little algebra, we can write the total Hamiltonian in the following form; 
	\br
	\label{hmm-cylinder}
	\hat H_1 &= &\frac{1}{2 a}\sum_{s=1}^{N_1}\sum_{m=-\infty}^{\infty}\sum_{j=1}^{N}\left[\hat\varXi^2 _{m,j,s}  +  (1+4 m^2) \frac{\hat\chi^2_{m,j,s}}{ j^2} + \frac{1}{s^2 \n^2}\hat\chi^2_{m,j,s} \r.\nn\\
	&& \l. 
	+  \l(\frac{\hat\chi_{m,j+1,s} -\hat\chi_{m,j-1,s}}{2}\right)^2  -\frac{\hat\chi_{m,j,s}}{2 j} \left(\hat\chi_{m,j+1,s} -\hat\chi_{m,j-1,s}\right) \r.\nn\\
	&&	\l. +\frac{\e}{P}\l\{\l(\hat\chi_{m,j+1,s}-2\hat\chi_{m,j,s}+\hat\chi_{m,j-1,s}\r)^2
	+\frac{2 \b}{j^2}\l(\hat\chi_{m,j+1,s}-2\hat\chi_{m,j,s}+\hat\chi_{m,j-1,s}\r)^2
	+\frac{\b^2}{j^4}\hat\chi_{m,j,s}^2\right\}\r.\nn\\
	&&\l. +\frac{\t}{P^2}\l\{\l(\frac{\hat\chi_{m,j+2,s}-2\hat\chi_{m,j+1,s}+2\hat\chi_{m,j-,s1}-\hat\chi_{m,j-2,s}}{2}\r)^2+\frac{\a}{j^2}\l(\hat\chi_{m,j+1,s}-2\hat\chi_{m,j,s}+\hat\chi_{m,j-1,s}\r)^2 \r.\r.\nn\\
	&&\l.\l. +\frac{\b^2}{j^4}\l(\frac{\hat\chi_{m,j+1,s}-\hat\chi_{m,j-1,s}}{2}\r)^2 
	+\frac{25/4+m^2}{j^6}\b^2\hat\chi_{m,j,s}^2-\frac{5\b^2 }{2j^5}\l(\frac{\hat\chi_{m,j+1,s}-\hat\chi_{m,j-1,s}}{2}\r)\hat\chi_{m,j,s}\r.\r.\nn\\
	&&\l.\l.-\l(\frac{\hat\chi_{m,j+2,s}-2\hat\chi_{m,j+1,s}+2\hat\chi_{m,j-1,s}-\hat\chi_{m,j-2,s}}{2j}\r)\l(\hat\chi_{m,j+1,s}-2\hat\chi_{m,j,s}+\hat\chi_{m,j-1,s}\r)\r.\r.\nn\\
	&&\l.\l.+ \frac{2\b}{j^2}\l(\frac{\hat\chi_{m,j+2,s}-2\hat\chi_{m,j+1,s}+2\hat\chi_{m,j-1,s}-\hat\chi_{m,j-2,s}}{2}\r)\l(\frac{\hat\chi_{m,j+1,s}-\hat\chi_{m,j-1,s}}{2}\r)\r.\r.\nn\\
	&&\l.\l.- \frac{5\b}{j^3}\l(\frac{\hat\chi_{m,j+2,s}-2\hat\chi_{m,j+1,s}+2\hat\chi_{m,j-1,s}-\hat\chi_{m,j-2,s}}{2}\r)\hat\chi_{m,j,s}-\frac{\b}{j^3}\l(\hat\chi_{m,j+1,s}-2\hat\chi_{m,j,s}+\hat\chi_{m,j-1,s}\r)\r.\r.\nn\\
	&&\l.\l.\times\l(\frac{\hat\chi_{m,j+1,s}-\hat\chi_{m,j-1,s}}{2}\r)
	+\frac{\l(5/2+2m^2\r)\b}{j^4}\l(\hat\chi_{m,j+1,s}-2\hat\chi_{m,j,s}+\hat\chi_{m,j-1,s}\r)\hat\chi_{m,j,s} \r\}\r]
	\er
	where  $\varXi_{m,j}=\sqrt{a}\;\Pi_{m,j}$ is the new canonical momentum field, $\chi_{m,j}=\vph_{m,j,p}/\sqrt{a}$ is the new dimensionless scalar field, $N_1$ and $N$ are the lattice points along the radial, $z$ directions respectively  and   $\chi_{m,N_1+1}=0,\chi_{m,N+1}=0, $ are the constraints on field.
	Thus the above Hamiltonian can be bring into a form of system of coupled HO's;
	\beq
	\label{hk1}
	H_1(P)=\frac{1}{2a}\sum_{s=1}^\infty\sum_{m=-\infty}^{N_1} \sum_{i,j=1}^N \l[\varXi_{m,s,i}^2\;\d_{i,j} +  \chi_{m,s,i}\; T_{ij}(P,m,s) \;\chi_{m,s,j}\r]
	\eeq  
	where $T_{ij}(P,m,s)=K_{ij}(P,m)+\dis\l(s\;\n\r)^{-2}\d_{i,j}$  is a real symmetric interaction  matrix have positive real energy eigenvalues. 
	Using the same procedure, we obtain ground state entanglement entropy. Fig.~\ref{Supp:cylinderentropy} contains the plot of entanglement 
	entropy versus area for scenario $\omega^2  = k^2 + k^4/\kappa^2$ and $\omega^2  = k^2 + k^4/\kappa^2+ k^6/\kappa^4$ respectively, where $k$ is the three dimensional wave vector. 
	\section{Specific heat  and Fidelity susceptibility from entanglement entropy}  
	R\'{e}nyi entropy is given by 
	\beq
	S_{\varsigma} = \frac{1}{1-\varsigma} \log \mbox{Tr}\; \rho^\varsigma
	\eeq	%\varsigma
	where $\varsigma$ is the R\'{e}nyi parameter and $\varsigma\rightarrow 1$ leads to von-Neumann entanglement entropy.  
	$ S_{\varsigma\rightarrow \infty} $ is referred in the literature as single copy entanglement entropy  \cite{eisert2005} and  quantifies quantum correlations in the quantum ground state of the 
	many body systems. The single copy entanglement entropy is given by, 
	\beq
	S_{m,s, \varsigma\rightarrow \infty}=-\sum_{i=1}^{N-n} \log{(1-\xi_{i})}
	\eeq
	where $\xi_{i}$  has the same structure as that in the circular case and is obtained from $T_{ij}(P,m,s)$. 
	
	It has been interpreted that there has been a close resemblance between the single copy entanglement entropy and 
	the thermodynamic entropy \cite{king2015, jean2012}. One can define entanglement specific heat as,  
	\beq
	C_{ent}=P\frac{dS_{\varsigma\rightarrow \infty}}{dP}
	\label{sp1}
	\eeq

	Fidelity susceptibility is defined as 
	\beq
	\chi_{F}= \dis\lim_{\d P\to 0}\l[-2 \;\frac{\d^2 \log F }{\d P^2}\r]
	\label{sp2}
	\eeq
	where $F=\langle\psi_{1\rm GS}(P+\d P)|\psi_{1\rm GS}(P)\rangle$ is the fidelity of the system and $ \psi_{1\rm GS}$ is the ground state wave function of the Hamiltonian $H_1$. 
	\subsection{ Implications for high temperature superconductivity }
	It is natural to ask: Is there any physical system that has a
	long-range interaction in the 2-D plane and shows such a
	distinct phase transition? One system is Copper Oxide HTS 
	which has 2-D layered crystal structure whose
	inter-atomic distance is smaller than the inter-atomic distance
	along the z-axis~ \cite{1994-Overend.etal-PRL}. It is
	long-known that, in HTS, the coloumbic interactions between
	the electrons tend to make an anti-ferromagnetic arrangement of
	spins in the Copper Oxide planes and the magnetic transition
	is controlled by the weak coupling between the planes along
	the $z$-axis \cite{Bonn2006}.
	\begin{figure}
		\centering	
		\includegraphics[scale=0.5]{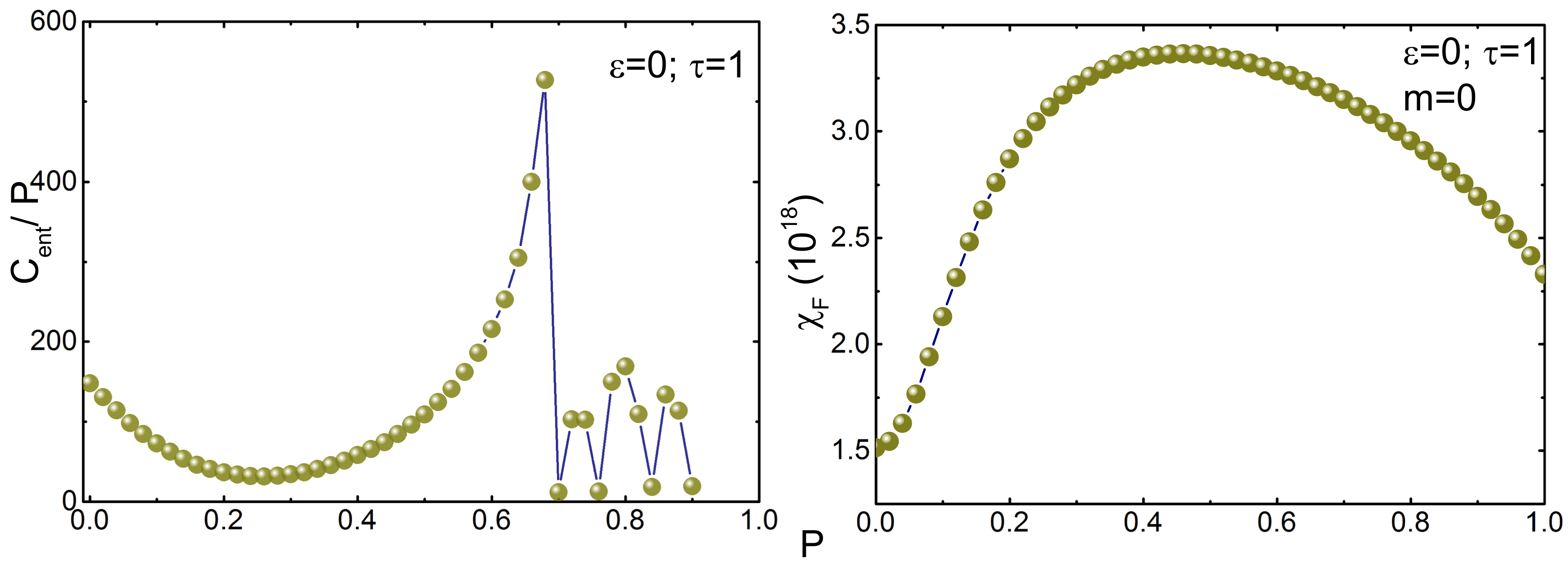}
		\caption{Left figure is the plot of entanglement
			specific heat per $P$ ($C_{ent}/P= d S_{\varsigma\rightarrow \infty}/d P $)
			versus $P$ obtained from the single copy
			entanglement entropy ($ S_\infty$ is the R\'{e}nyi
			entropy having infinity as the R\'{e}nyi
			index). Right figure is the plot of fidelity
			susceptibility ($\chi_F= \dis\lim_{\d P\to 0}-2 \log
			\d^2 F/\d P^2$) as a function of $P$. We have taken
			lattice size of the $z$-axis to be 10 times more
			than the lattice size of the 2-dimensional surface,
			$N = 600, n=300$ and $\d P=10^{-7}$. }
		%\label{Fig:entangspheat}
	\end{figure}
	To overcome the complexity of the interactions, we consider a
	scalar order parameter $\Phi$ in 3-dimensional cylindrical
	geometry such that the higher derivative terms contribute only
	in the 2-dimensional plane while the first derivative term
	contribute acts in all the three spatial dimensions. Repeating
	the analysis in 3-dimensions, it can be shown that the model in
	3-dimensions has the same entropy profile as that in the
	2-dimensional case. In Fig.~(\ref{Fig:entangspheat}), we have
	plotted entanglement specific heat and fidelity susceptibility  defined in Eqs. (\ref{sp1}) and (\ref{sp2})
	as a function of $P$.  
	
	Following points are interesting to
	note: First, the entanglement specific heat shows discontinuity at a
	particular value of $P$. This is indeed similar to the
	discontinuity of the specific heat measurement of the single
	crystals of YBa$_2$Cu$_3$O$_{7-\delta}$~\cite{1994-Overend.etal-PRL}. 
	Second, discontinuity of the entanglement specific heat at, say, $P_0$ 
	implies that the correlation length diverges close to $P_0$. For instance, 
	in the case of transverse quantum Ising model, entanglement specific heat diverges 
	logarithmically \cite{Osterloh2002} that signals the correlation length to 
	diverge. In our case, the entanglement specific heat diverges more like power-law. 
	Third, it is interesting to see how our model fares with the
	specific heat measurements at extreme high magnetic fields as
	recently reported by Badoux et al \cite{Badoux2016}.             
%\bibliographystyle{plainnat.bst}
%	\bibliographystyle{utphys.bst}
%\bibliography{library.bib}   
\providecommand{\href}[2]{#2}\begingroup\raggedright\endgroup

\end{document}